\documentclass[preprint2]{aastex}
\usepackage{graphicx}
\usepackage{natbib}
\usepackage{enumerate}
\usepackage{amssymb}
\usepackage{amsmath}

\shorttitle{}
\shortauthors{}

\begin{document}


\title{THE COMPLEXITY THAT THE FIRST STARS BROUGHT TO THE UNIVERSE: FRAGILITY OF METAL ENRICHED GAS IN A RADIATION FIELD}


\author{A. Aykutalp\altaffilmark{1} and M. Spaans\altaffilmark{1}}
\affil{Kapteyn Astronomical Institute, University of Groningen, PO Box 800, 9700 AV Groningen, The Netherlands}

\email{aykutalp@astro.rug.nl}
\email{spaans@astro.rug.nl}



\begin{abstract} The initial mass function (IMF) of the first (Population III) stars and Population II (Pop II) stars is poorly known due to a lack of observations of the period between recombination and reionization. In simulations of the formation of the first stars, it has been shown that, due to the limited  ability of metal-free primordial gas to cool,  the IMF of the first stars is a few orders of magnitude more massive than the current IMF. The transition from a high-mass IMF of the first stars to a lower-mass current IMF is thus important to understand. To study the underlying physics of this transition, we performed several simulations using the cosmological hydrodynamic adaptive mesh refinement code Enzo for metallicities of 10$^{-4}$, 10$^{-3}$, 10$^{-2}$, and 10$^{-1}$ $Z_{\odot}$. In our simulations, we include a star formation prescription that is derived from a metallicity dependent multi-phase interstellar medium (ISM) structure, an external UV radiation field, and a mechanical feedback algorithm. We also implement cosmic ray heating, photoelectric heating, and gas$-$dust heating/cooling, and follow the metal enrichment of the ISM. It is found that the interplay between metallicity and UV radiation leads to the coexistence of Pop III and Pop II star formation in non-zero metallicity ($Z/Z_{\odot}$ $\geq$10$^{-2}$) gas. A cold ($T < $100 K) and dense ($\rho$$>$10$^{-22}$ g cm$^{-3}$) gas phase is fragile to ambient UV radiation. In a metal-poor ($Z/Z_{\odot}$ $\leq$10$^{-3}$) gas, the cold and dense gas phase does not form in the presence of a radiation field of $F_{0}$$\sim$10$^{-5}$$-$10$^{-4}$ erg cm$^{-2}$ s$^{-1}$. Therefore, metallicity by itself is not a good indicator of the Pop III$-$Pop II transition. Metal-rich ($Z/Z_{\odot}$$\geq$10$^{-2}$) gas dynamically evolves two to three orders of magnitude faster than metal-poor gas ($Z/Z_{\odot}$$\leq$10$^{-3}$). The simulations including supernova explosions show that pre-enrichment of the halo does not affect the mixing of metals.\\
\end{abstract}




\section{INTRODUCTION} In the last two decades, cosmological simulations have become an important tool for theoreticians to simulate the structure formation in the universe from the primordial density fluctuations. The cosmological model that currently matches the observations of the cosmic microwave background (CMB) best is the so-called cold dark matter (CDM) model. According to hierarchical structure formation \citep{1974ApJ...187..425P}, small clumps are the first to collapse and form structures, and then through mergers and the accretion of matter they build up larger structures. Cosmological simulations, based on the CDM model of hierarchical structure formation, predict that the first stars  (30$-$300 $M_{\odot}$) have formed at redshifts $z \sim$ 20-30, in dark matter halos with masses of $\sim$ 10$^6$ $M_{\odot}$ \citep{1997ApJ...474....1T, 2002Sci...295...93A, 2002ApJ...564...23B, 2003ApJ...592..645Y, 2007ApJ...654...66O}. \\
\indent In the current literature, the first stars are called Population III (Pop III) stars. The metallicity of Pop III stars is so low that metal cooling does not have any effect on their formation. In the simulations of the formation of Pop III stars, it has been shown that, due to the limited ability of metal-free primordial gas to cool, the initial mass function (IMF) of Pop III stars is a few orders of magnitude more massive than the current IMF \citep{2002Sci...295...93A, 2006ApJ...652....6Y, 2002ApJ...564...23B}. More recently, it has been suggested that Pop III stars are not necessarily that massive and can be in the range 10$-$100 $M_{\odot}$ \citep{2006ApJ...641....1T, 2007ApJ...665.1361T, 2007ApJ...664L..63T}. On the other hand, from observations of the nearby universe we know that the present day stellar mass scale is $\sim$ 0.3 $M_{\odot}$ \citep{2002Sci...295...82K, 2003PASP..115..763C}.\\
\indent The chemistry of  zero metallicity (primordial) gas has been studied by a number of authors \citep{1987IAUS..120..109D, 1997NewA....2..181A, 1998A&A...335..403G, 2001ASPC..222..129A, 2008AIPC..990...25G, 2008MNRAS.388.1627G, 2011ApJ...726...55T}. Once the gas has virialized in the potential wells of dark matter halos, additional cooling is required for the further collapse of the gas and to form stars. The formation of a star depends on the ability of interstellar gas to cool and form dense molecular clouds. The modest cooling ability of primordial gas leads to high masses for Pop III stars. The cooling efficiency of star forming gas will be significantly affected by the addition of radiation and metals after the formation of the first stars. Hence, the chemical composition and the radiation environment of interstellar gas are the key parameters to study.\\
\indent The main coolants for primordial gas with temperatures $T \geq$ 10$^4$ K are Ly$\alpha$ emission of neutral atomic hydrogen (H I, 1216 \AA), and ionized helium (He II, 304 \AA). Below this temperature, the dominant coolant in primordial gas is molecular hydrogen (H$_2$). \cite{1967Natur.216..976S} realized the importance of gas phase H$_2$ formation in a primordial gas for the formation of protogalactic objects. At low densities ($n <$ 10$^8$ cm$^{-3}$), H$_2$ can form via intermediate ions H${_2}^{+}$ and H$^-$. The cooling rate of H$_2$ ($\Lambda\rm{_{H_2}}$) scales with density as $\Lambda\rm{_{H_2}} \propto n^2$ at low densities ($n < 10^4 cm^{-3}$), where radiative de-excitation dominates, and as $\Lambda\rm{_{H_2}} \propto n$ at high densities ($n >$ 10$^4$ cm$^{-3}$), where collisional de-excitation dominates. Molecular hydrogen does not have a permanent electric dipole moment while the hydrogen deuteride (HD) molecule does, which makes HD a better coolant than H$_2$. Although the primordial deuterium abundance is small relative to hydrogen ($n\rm{_D}$$/$$n\rm{_H}$ = 4$\times$10$^{-5}$), chemical fractionation leads to an enhancement of the ratio, $n\rm{_{HD}}$$/$$n\rm{_{H_2}}$ $\approx$ 10$^{-3}$ (\cite{1993A&A...267..337P, 1998A&A...335..403G, 1998ApJ...509....1S}). If primordial gas is significantly ionized then HD cooling can lower the temperature to the level of the CMB at $z \sim$ 10$-$20 \citep{2006MNRAS.366..247J}.\\
\indent When the first stars form, the universe becomes much more complex. According to their initial masses, they will either explode as core-collapse supernovae (10 $M_{\odot}$ $<$ $M_*$ $<$ 140 $M_{\odot}$) or as pair instability supernovae  (PISN; 140 $M_{\odot}$ $<$ $M_*$ $<$ 260 $M_{\odot}$,  \cite{2002ApJ...567..532H}). Through these supernova explosions (SNe), the interstellar medium (ISM), and the intergalactic medium (IGM) will be enriched with metals. From that point on one has to take into account the cooling from the fine-structure lines of metals and the rotational transitions of molecules like CO. Moreover, gas$-$dust heating/cooling can be important as well, in the sense that dust grains will be heated to at least the CMB temperature. Dust allows the efficient formation of H$_2$ and HD \citep{2004ApJ...611...40C, 2009A&A...496..365C}, and may collisionally heat/cool the gas depending on the sign of gas temperature minus dust temperature \citep{2002ApJ...571...30S}. Also, dust grains attenuate UV radiation. For $z >$ 10, CMB photons can be an excitation source (radiative pumping) of atomic and molecular levels \citep{2009ApJ...691..441S}. This allows the CMB temperature to act as a thermodynamic floor, below which gas cannot cool, provided that collisional de-excitation dominates the removal of population from excited states (\cite{2000ApJ...538..115S, 2005ApJ...626..644S}). Furthermore, SNe will initiate shock waves that propagate through the ISM. These shock waves can heat up the ISM and cause a delay in the formation of the next generation of stars or can compress the gas, which makes it collapse and hence give rise to further star formation \citep{2003ApJ...596L.135B, 2003MNRAS.339..289S, 2008ApJ...682...49W}  .\\ 
\indent The aim of this work is to (1) compute at what metallicity a cold and dense gas phase emerges and (2)  assess the sensitivity of this phase to background UV radiation. This paper is structured as follows. In Section \ref{Sim}, we detail our cosmological/hydrodynamic simulation. Specifically, we discuss the relevant cooling and heating processes for star formation and the global characteristics of early star formation in primordial galaxies. In Section \ref{Res}, we highlight the implications of our findings. Finally, in Section \ref{Fin} we discuss our results and present our conclusions.\\
\section{SIMULATIONS}\label{Sim} In this work, we use the cosmological adaptive mesh refinement code Enzo \citep{1997ASPC..123..363B, 2004astro.ph..3044O}. We perform simulations in a three-dimensional periodic box with a side length of 1 $h^{-1} Mpc$, initialized at $z =$ 99. The size of the root grid is 128$^3$ with three nested subgrids, each refined by a factor of two. The finest grid has an effective resolution of 1024$^3$  with a side length of 125 $h^{-1} kpc$. This resolution results in a dark matter and baryonic matter resolution of 2 and 0.4 $M_\odot$, respectively.  Refinement is restricted to the finest grid and occurs during the simulations whenever the baryonic matter, or dark matter density, is greater than the mean density by a factor of four or eight, respectively. The maximum level of refinement that is reached in the finest grid is eight. Refinement occurs such that the Jean length is always resolved by at least eight cells, this ensures that we meet the Truelove criterion, which requires the Jeans length to be resolved by at least four cells on each axis \citep{1997ApJ...489L.179T}. The virial mass of our progenitor halo at redshift $z =$ 21 is $M_{vir}$ =  3.4$\times$10$^6$ $M_\odot$, where $M_{vir}$ is the mass in a sphere that encloses an average dark matter overdensity of 200. We use Wilkinson Microwave Anisotropy Probe five-year cosmological parameters \citep{2009ApJS..180..330K}, which have the following values: $\Omega _{\Lambda}$ = 0.7208, $\Omega _{m}$ = 0.233,  $ \Omega _b$ = 0.0462, $\sigma_8$ = 0.9, and $h =$ 0.701. Here, $\Omega _{\Lambda}$ is the vacuum energy, $\Omega _{m}$ is the matter density, $\Omega _b$ is the baryon density, $\sigma_8$ is the variance of random mass fluctuations in a sphere of radius 8 $h^{-1}$ Mpc, and $h$ is the Hubble parameter in units of 100 km s$^{-1}$ Mpc$^{-1}$. We focus on a single halo with a dark matter mass of $\sim$ $10^9$ $M_\odot$ at $z \sim$ 5 which is expected to be a typical mass for the halo population at that redshift.\\
\indent  For the analysis of our cosmological simulations we use YT, a cross-platform analysis toolkit written in Python\citep{SciPyProceedings_46, 2011ApJS..192....9T}.\\
\subsection{Cooling and Heating Processes in the ISM}\label{CoHe} \indent Metal enriched gas cools more efficiently by fine-structure lines of [C II] (157.74 $\mu$m), [O I] (63.18 $\mu$m, 145.5 $\mu$m), [Si II] (34.8 $\mu$m), [Fe II] (25.99 $\mu$m, 35.35 $\mu$m), and rotational lines of CO than by HD or H$_2$ emission \citep{2006ApJ...643...26S}. In the outermost layers of a star-forming cloud, the so-called photon dominated region (PDR), temperatures can increase up to 1000 K due to a strong UV radiation field, and cooling results from the fine-structure lines of [C II] and [O I]. Deeper into the cloud, the temperature decreases to $\sim$ 30 K through the balance between cosmic ray and dust heating and low-$J$ CO rotational line cooling.\\
\indent  In our simulations, the gas is homogeneously pre-enriched to some non-zero metallicity at redshift of $z =$ 30. We perform a series of simulations using cooling models for four different metallicities, 10$^{-4}$, 10$^{-3}$, 10$^{-2}$, and 10$^{-1}$ $Z_{\odot}$, derived from the chemical network of \cite{2005A&A...436..397M}. In their PDR code, they include cooling from fine-structure lines of carbon (C$^+$ and C), oxygen (O), molecular lines from species like carbon monoxide (CO), H$_2$, HD, and water (H$_2$O). All level populations are computed under statistical equilibrium and the chemistry includes gas phase and grain surface formation of H$_2$ and HD \citep{2004ApJ...611...40C, 2009A&A...496..365C} and line trapping using the multi-zone escape probability method of \cite{2005A&A...440..559P}. The cooling tables that we use, depend on FUV radiation field strength, ambient gas velocity dispersion, temperature, metallicity, and H$_2$ abundance. As such, they provide an accurate treatment of the thermal and chemical balance of low-metallicity PDRs. The radiation field strength ($G_0$) here is the flux between 6$-$100 eV and it extends beyond 13.6 eV according to a starburst99 spectrum for a Salpeter IMF, i.e., the H II region is computed as part of the PDR. Typically, the UV flux diminishes very quickly beyond 13.6 eV. The cooling tables enjoy a range in irradiation ($G_{0}$=10$^{-2}$, 10$^{-1}$, 10$^{0}$, 10$^{1}$, 10$^{2}$, 10$^{3}$, 10$^{4}$),  H$_2$ abundance ($f_{H_2}$= 10$^{-5}$, 10$^{-4}$, 10$^{-3}$, 10$^{-2}$, 10$^{-1}$, 0.2, 0.4, 0.5) and metallicity (10$^{-4}$, 10$^{-3}$, 10$^{-2}$, 10$^{-1}$, 1 $Z_{\odot}$). We adopt Milky Way like abundance ratios based on the values of \cite{2005ASPC..336...25A} and \cite{2004oee..symp..336J}. A summary of our simulations is listed in Table \ref{tb:liste}.\\
\indent  At high column densities cooling and heating can be suppressed due to optical depth and dust opacity effects. In the case of an optically thick medium, a large part of the emitted photons are thus reabsorbed. This results in a lower critical density where the critical density is defined as the density at which the radiative de-excitation rate equals the collisional de-excitation rate. At densities much below the critical density ($n \ll n_{cr}$), radiative de-excitation dominates over collisional de-excitation, whereas at high densities ($n > n_{cr}$), collisions dominate the de-excitation process and the gas is in local thermodynamic equilibrium (LTE).  Radiation is trapped if the opacity is concentrated on small physical scales which is the case in our simulations for the warm and dense gas. Since we do not resolve these PDR sub-structures in our simulations we approximate these opacity effects as follows.\\
\indent We adopt  a turbulent coherence length $L$ of 0.3 pc and a local turbulent velocity dispersion of $dV =$ 3 km s$^{-1}$ to mimic the properties of a turbulent region. This yields a formal velocity gradient $dV/L$ for scales larger than $L$. When the cooling is either optically thin ($\tau$ $\ll$ 1) or optically thick ($\tau$ $\gg$ 1) the choice of $dV$ and $L$ is not important. For the radiative transfer in the cooling lines (under statistical equilibrium), we use the multi-zone escape probability code of \cite{2005A&A...440..559P} . Each zone in a PDR is treated this way, and a correction to the cooling and heating rates due to line trapping and optical depth effects is derived. See \cite{2009ApJ...702...63W} for more details.\\
\indent For an incident UV radiation field, photo-electric emission from (small) dust grains and polycyclic aromatic hydrocarbons is the dominant heating source  in the neutral ISM. Dust grains can absorb an FUV photon which leads to the ejection of an electron, carrying some of the photon energy away in the form of kinetic energy. Then, through elastic collisions this excess kinetic energy heats up the gas. In our simulations we add photoelectric heating through
\begin{equation}
\Gamma \rm{_{grain}} = 10^{-24} \epsilon G^{'}_{0,dust} n_H \: erg \: cm^{-3}\: s^{-1}.
\end{equation}
Here, $G^{'}_{0,dust}$ is the radiation field attenuated by dust absorption and is given by
\begin{equation}
G^{'}_{0,dust} = G_0 \exp(-1.8A_{\nu}) 
\end{equation}
where $G_0 =$ 1, in Habing units, corresponds to a flux of 1.6$\times$10$^{-3}$ erg cm$^{-2}$s$^{-1}$, $A_{\nu}$ is the line-of-sight visual extinction at optical wavelengths caused by interstellar dust  and the total number density of hydrogen is, $n\rm{_H}$ = $n$(H) + 2$n$(H$_2$) \citep{ 1994ApJ...427..822B, 2005A&A...436..397M}. $\epsilon$ is the heating efficiency, dependent on $G_0$, the kinetic gas temperature $T_k$, and the electron density $n_e$ as $G_0$$T^{1/2}_k$/$n_e$, and is given by
\begin{equation}
\begin{split}
\epsilon = \frac{4.87 \times 10^{-2}}{[1+4 \times 10^{-3}(\frac{G_{0}T^{1/2}_{k}}{n_{e}})^{0.73}]} \\  
 +\frac{3.65\times10^{-2}(\frac{T_{k}}{10^{4}})^{0.7}}{[1+2\times10^{-4}(\frac{G_{0}T^{1/2}_{k}}{n_{e}})]}.
\end{split}
\end{equation}
\indent On the other hand, gas and dust temperatures are not equal in the ISM. At high number densities, $n_H$ $>$10 $^{4.5}$ $(\frac{Z}{Z_{\odot}})^{-1}$, gas$-$dust heating/cooling becomes important and therefore we add gas$-$dust heating/cooling into our simulations as given by \citep{2005A&A...436..397M};
\begin{equation}
\begin{split}
\Gamma\rm{_{coll}} = 1.2\times10^{-31}n_{H}^{2}(\frac{T_{k}}{100})^{1/2} (\frac{1000 \AA}{a_{min}})^{1/2} \\
\times [1-0.8 \exp(\frac{-75}{T_{k}})](T_{d}-T_{k}),
\end{split}
\end{equation}
where $T_d$ is the dust temperature, $T_k$ is the gas temperature, and $a_{min}$ is the minimum grain size which we take as 1 $\mu$m \citep{2003ApJ...598..785N}.\\
\indent Deeper into a cloud the UV radiation is attenuated by dust and cosmic ray heating becomes important. Ionization by cosmic rays produces energetic electrons. \cite{1973ApJ...179L.147G} and \cite{1978ApJ...219..750C} calculated that about 8 eV  of heat is deposited in a molecular gas per primary ionization. Including helium ionization, \cite{1985ApJ...291..722T} find for the heating rate
\begin{equation}
\Gamma\rm{_{CR}} = 1.5 \times 10^{-11} \zeta n(H_2) \: erg \: cm^{-3}\: s^{-1},
\end{equation}
where $\zeta$ is the cosmic ray ionization rate per H$_2$ molecule. In our simulations, we scale $\zeta$ with the derived star formation rate (SFR) of our simulations such that for an SFR of 1 $M_{\odot}$ yr$^{-1}$ and a Salpeter-like IMF between 1 and 100 $M{\odot}$, we take $\zeta$ to be 3$\times$10$^{-17}$s$^{-1}$ \citep{2005ApJ...626..644S}.
\begin{deluxetable}{@{}lccccc}
\tabletypesize{\scriptsize}  
\tablecaption{Simulation parameters}              
\tablewidth{0pt}
\tablecolumns{5}            
\tablehead{ \colhead{Run}&\colhead{$L$}&\colhead{Metallicity}&
\colhead{SFFB}&\colhead{ G$_0$}\\
& (Mpc) & ($Z_{\odot}$) & & (Habing)} \\  
\startdata
Z1-G1 & 1 & 10$^{-1}$ & No & 10$^{-2}$ \\
Z2-G1 & 1 & 10$^{-2}$ & No & 10$^{-2}$ \\
Z3-G1 & 1 &10$^{-3}$ & No & 10$^{-2}$ \\
Z4-G1 & 1 & 10$^{-4}$ & No & 10$^{-2}$ \\
Z1-G10 & 1 & 10$^{-1}$ & No & 10$^{-1}$ \\
Z2-G10 & 1 & 10$^{-2}$ & No & 10$^{-1}$ \\
Z3-G10 & 1 & 10$^{-3}$ & No & 10$^{-1}$ \\
Z4-G10 & 1 & 10$^{-3}$ & No & 10$^{-1}$ \\
ZS2 & 1 & 10$^{-2}$ & Yes & 10$^{-2}$\\ 
ZS4 & 1 & 10$^{-4}$ & Yes & 10$^{-2}$ \\
\enddata
\label{tb:liste}
\tablecomments{Column 1: Simulation name; Column 2: Simulation box size; Column 3:  Simulation pre-enrichment; Column 4: Star formation and feedback; Column 5: UV background strength.}
\end{deluxetable}
\subsection{Star Formation}\label{St} In order to understand the underlying physics of the transition from a high-mass IMF of Pop III stars to a lower mass current IMF we have run simulations for different metallicities and different UV radiation backgrounds and have incorporated the metal yields of SNe.\\
\indent In our simulations where star formation is included, a star particle is formed when a grid cell has a density exceeding 0.01 cm$^{-3}$ and a temperature below 10$^4$ K. We follow the hydrodynamic transport of metals produced by the SNe to the enrichment of halos. We do not use the chemical network of Enzo, except for the non-equilibrium formation of $H_2$ (rate
equation method of \cite{1997NewA....2..209A}) but our $\Lambda$($G_0$, $f_{H_2}$, Z/$Z_{\odot}$) cooling tables instead.\\ 
\indent In order to compute the metal yield by the first stars we need to decide on some parameters like the mass of the halo, IMF of the first stars, star formation efficiency, and the number of SNe. From their simulations of pre-galactic structure formation, \cite{2001ApJ...548..509M} determined that the minimum mass of a halo that hosts a massive primordial star is
\begin{equation}
\begin{split}
\frac {M_{min}}{{M_\odot}} = \exp(\frac{f_{cd}}{0.06})(1.25\times10^5 \\
+ 8.7 \times10^5 F^{0.47}_{LW,-21}),
\end{split}
\end{equation}
where $M_{min}$ is the minimum halo mass that contains a cold dense core; $f_{cd}$ is the fraction of gas that is cold ($T <$ 0.5 $T_{vir}$) and dense ($\rm{\rho}$ $>$ 330 cm$^{-3}$);,and $F_{LW}$ is the flux within the Lyman$-$Werner (LW) bands in units of 10$^{-21}$ erg s$^{-1}$ cm$^{-2}$ Hz$^{-1}$. We choose $f_{cd}$ = 0.02 for a typical UV background of $J =$ 10$^{-21}$  erg s$^{-1}$ cm$^{-2}$ Hz$^{-1}$ sr$^{-1}$, so that a halo with a minimum mass of 4.16$\times$10$^6$ $M_{\odot}$ can support cold dense gas which will eventually form a primordial star. Note here that $J$ = $J_{-21}$ corresponds to a total UV flux of $F_0$ $\simeq$ 4$\times$$10^{-5}$  erg s$^{-1}$ cm$^{-2}$ or $G_0$ $\simeq$ 2.5$\times$$10^{-2}$. For a Salpeter like IMF, $G_0$ $\simeq$ 1 corresponds to an SFR of about 3 $M_{\odot}$ yr$^{-1}$ (10 kpc)$^{-2}$.\\
\indent High redshift, low mass galaxies are conjectured to have properties similar to those of local dwarf galaxies. In these galaxies the star formation efficiency, $\rm{f_{\star}}$, ranges from 0.02 to 0.08 \citep{1999A&A...349..424T}. In this work we take  $f_{\star} =$ 0.05.\\
\section{RESULTS}\label{Res} 
\subsection{Multi-phase ISM} 
\begin{figure*}
\includegraphics[angle=0,width=8cm]{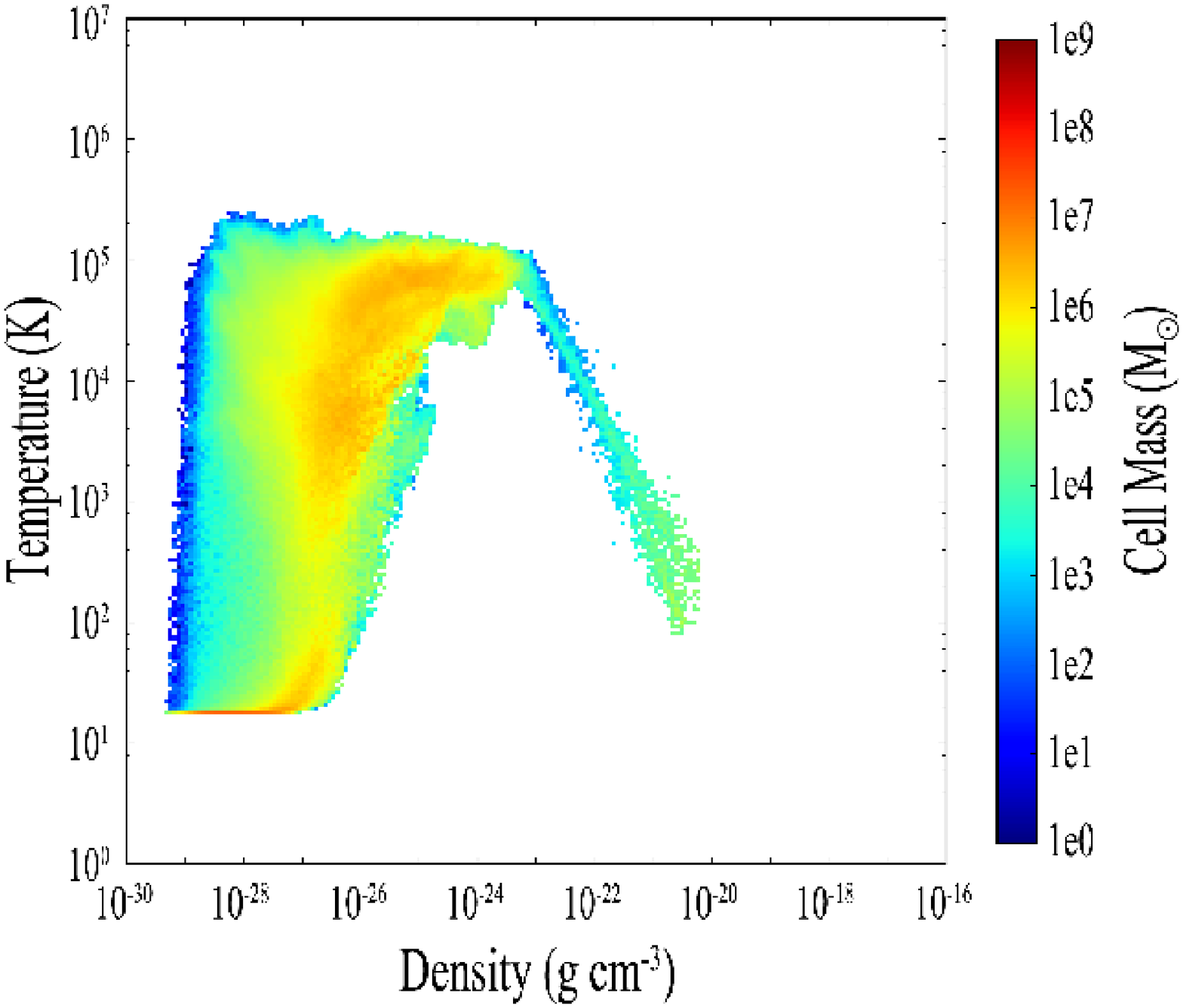}
\hspace{0.5cm}
\includegraphics[angle=0,width=8cm]{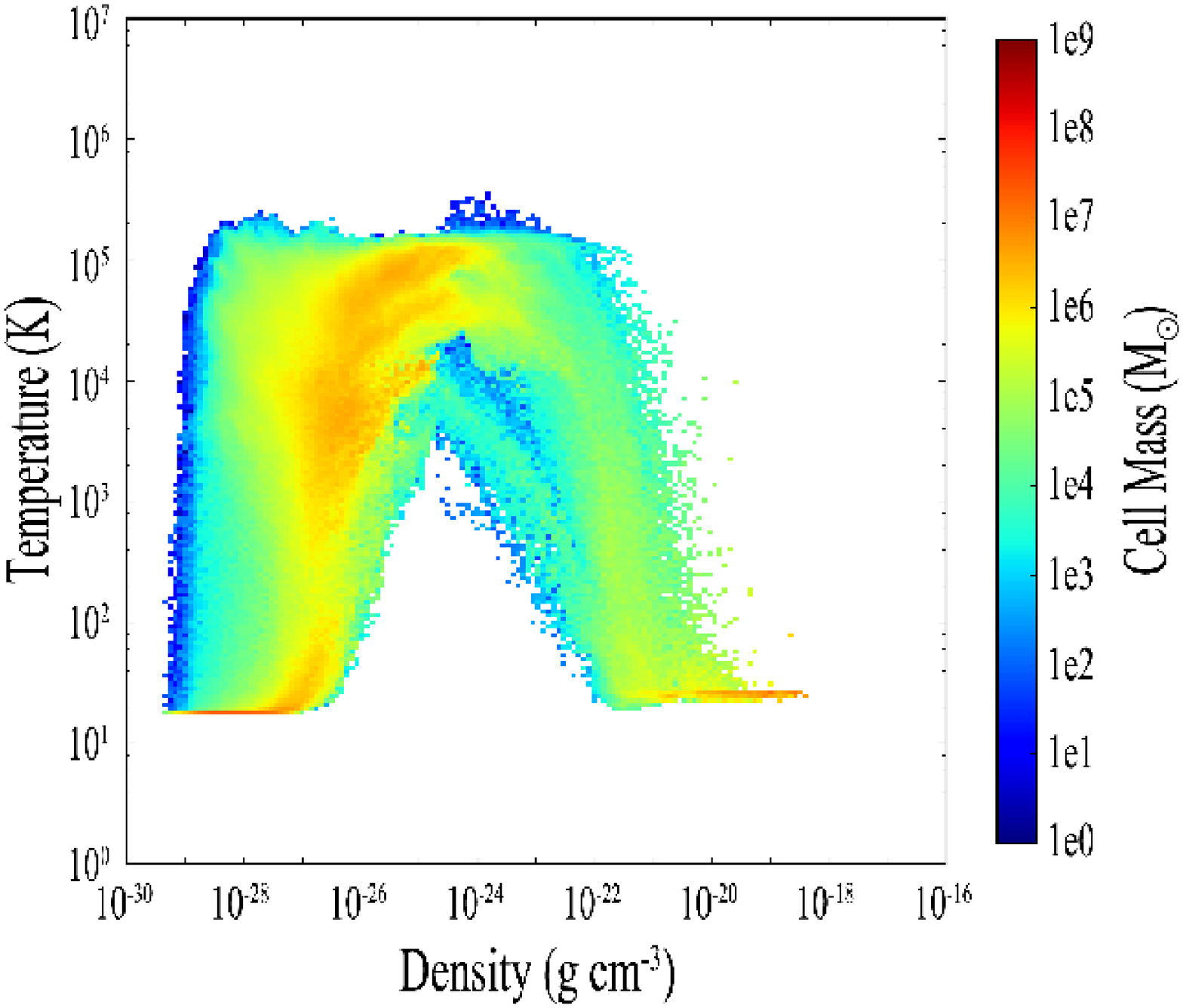}
\caption{Density$-$temperature profile of the central 50 kpc of a halo at redshift $z \sim$ 5 for metallicities of $Z/Z_{\odot}$=10$^{-4}$ (left) and $Z/Z_{\odot}$=10$^{-2}$ (right) for a radiation field of $G_0$=10$^{-2}$.}\label{fig1}
\end{figure*}

\begin{figure*}
\includegraphics[angle=0,width=8cm]{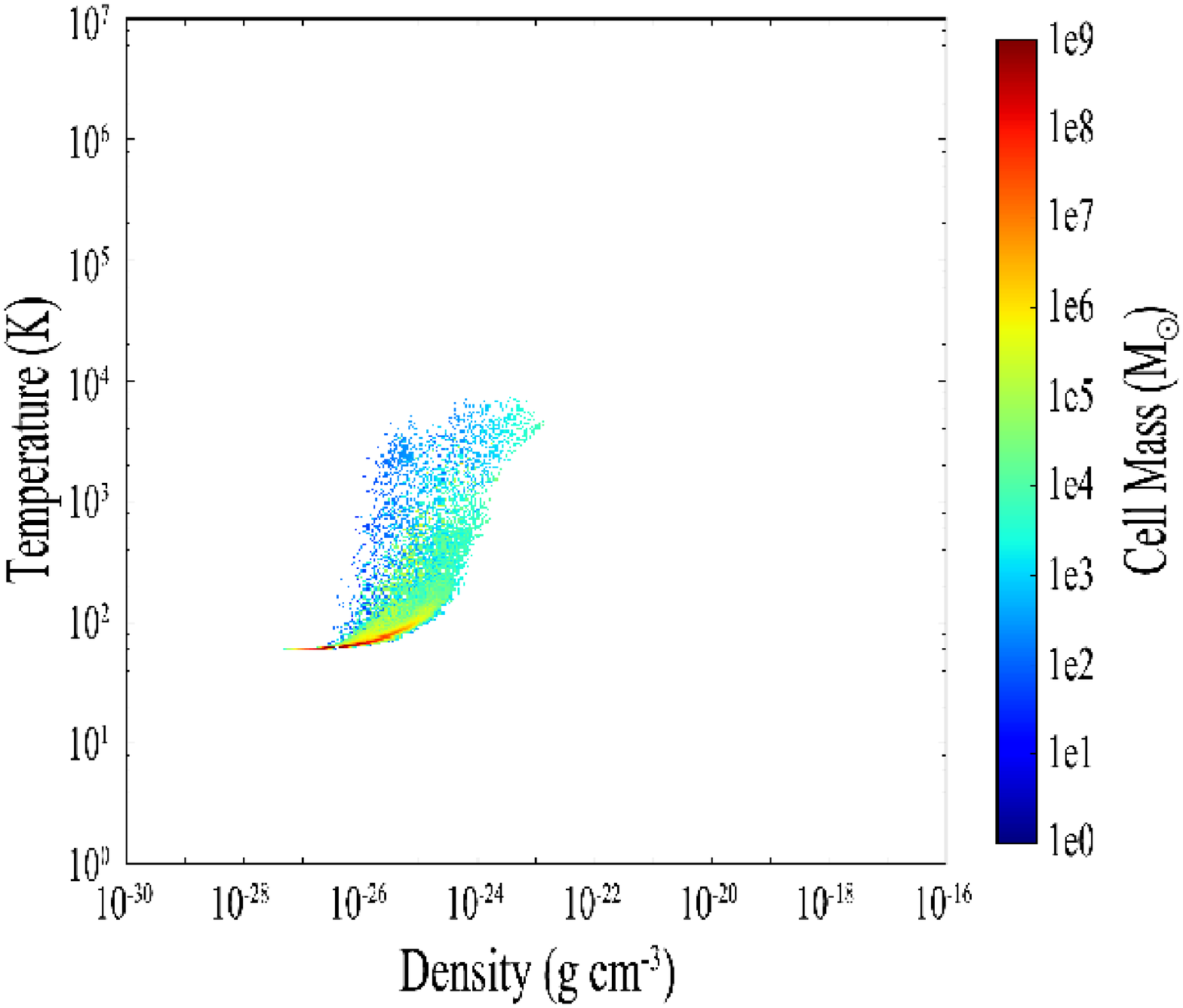}
\hspace{0.5cm}
\includegraphics[angle=0,width=8cm]{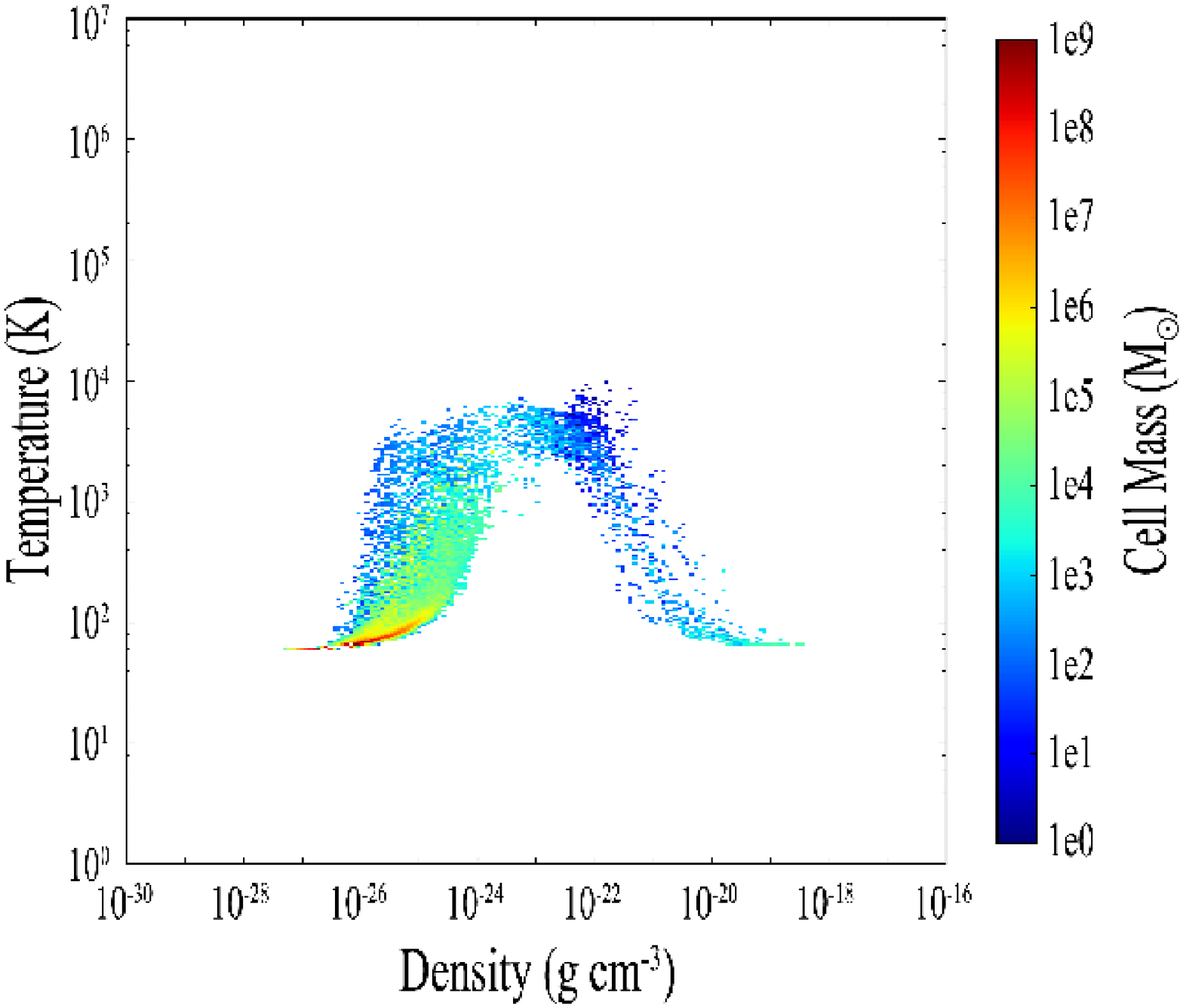}
\caption{Density$-$temperature profile of the central 50 kpc of a halo at redshift $z \sim$ 20 for metallicities of $Z/Z_{\odot}$=10$^{-4}$ (left) and $Z/Z_{\odot}$=10$^{-2}$ (right) for a radiation field of $G_0$=10$^{-2}$.}\label{fig2}
\end{figure*}

\begin{figure*}
\includegraphics[angle=0,width=8.3cm]{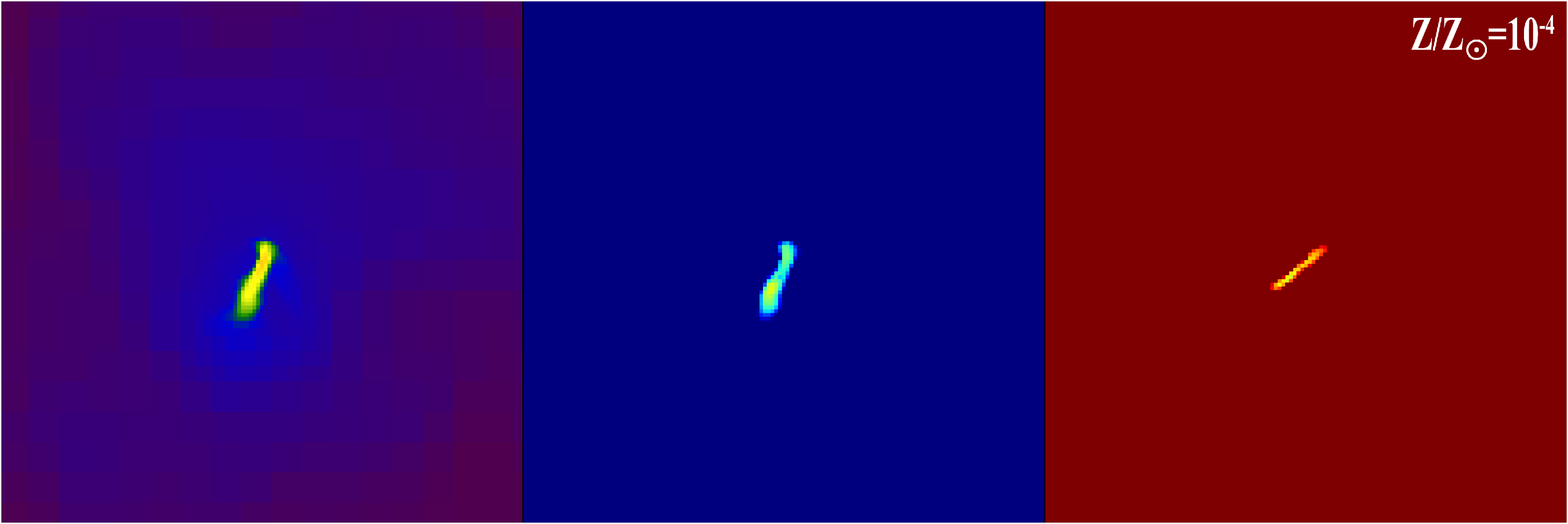}
\vspace{0.05cm}
\includegraphics[angle=0,width=8.3cm]{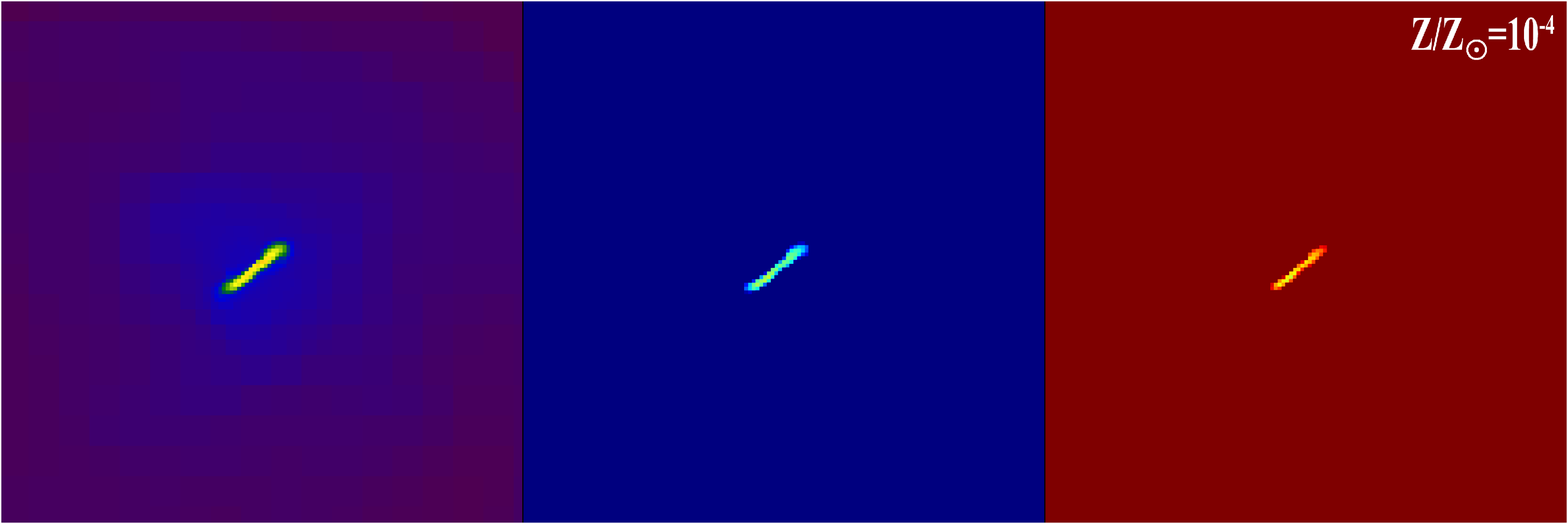}\\
\includegraphics[angle=0,width=8.3cm]{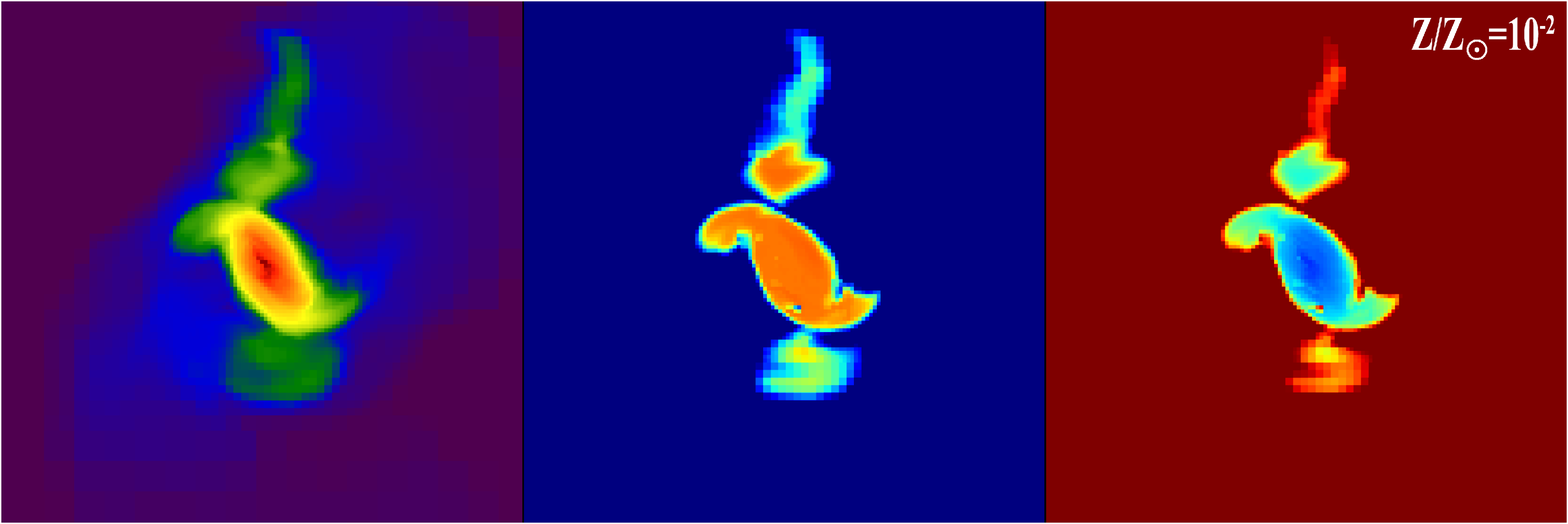}
\vspace{0.05cm}
\includegraphics[angle=0,width=8.3cm]{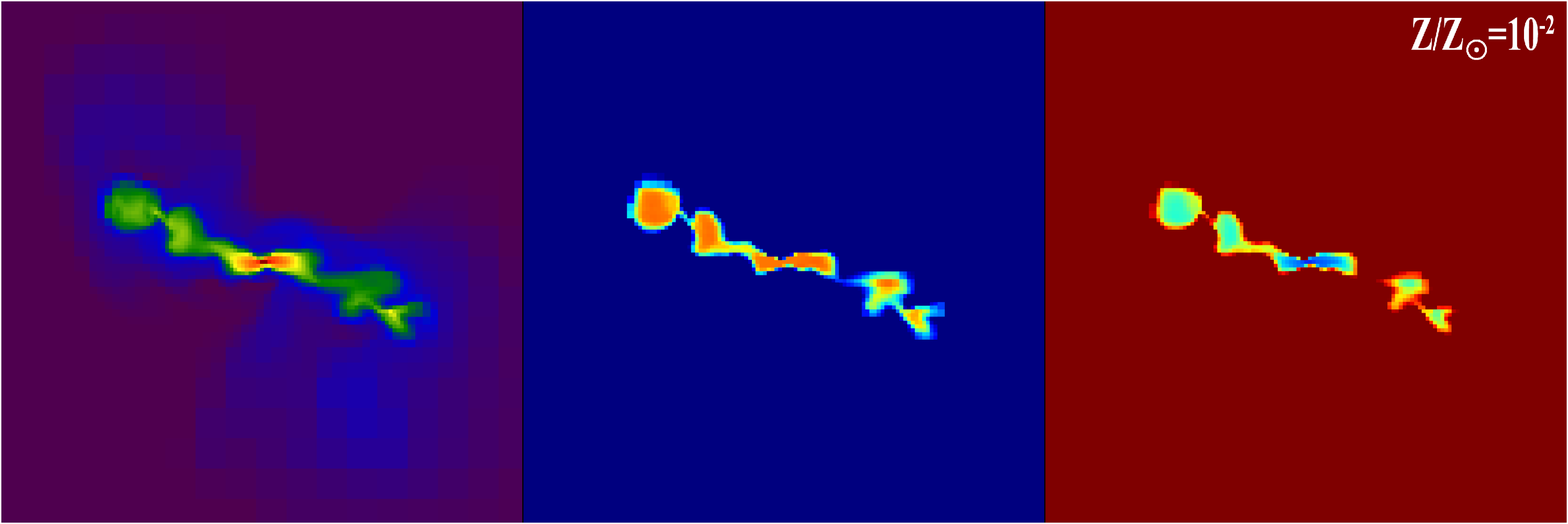}\\
\includegraphics[angle=0,width=8.3cm]{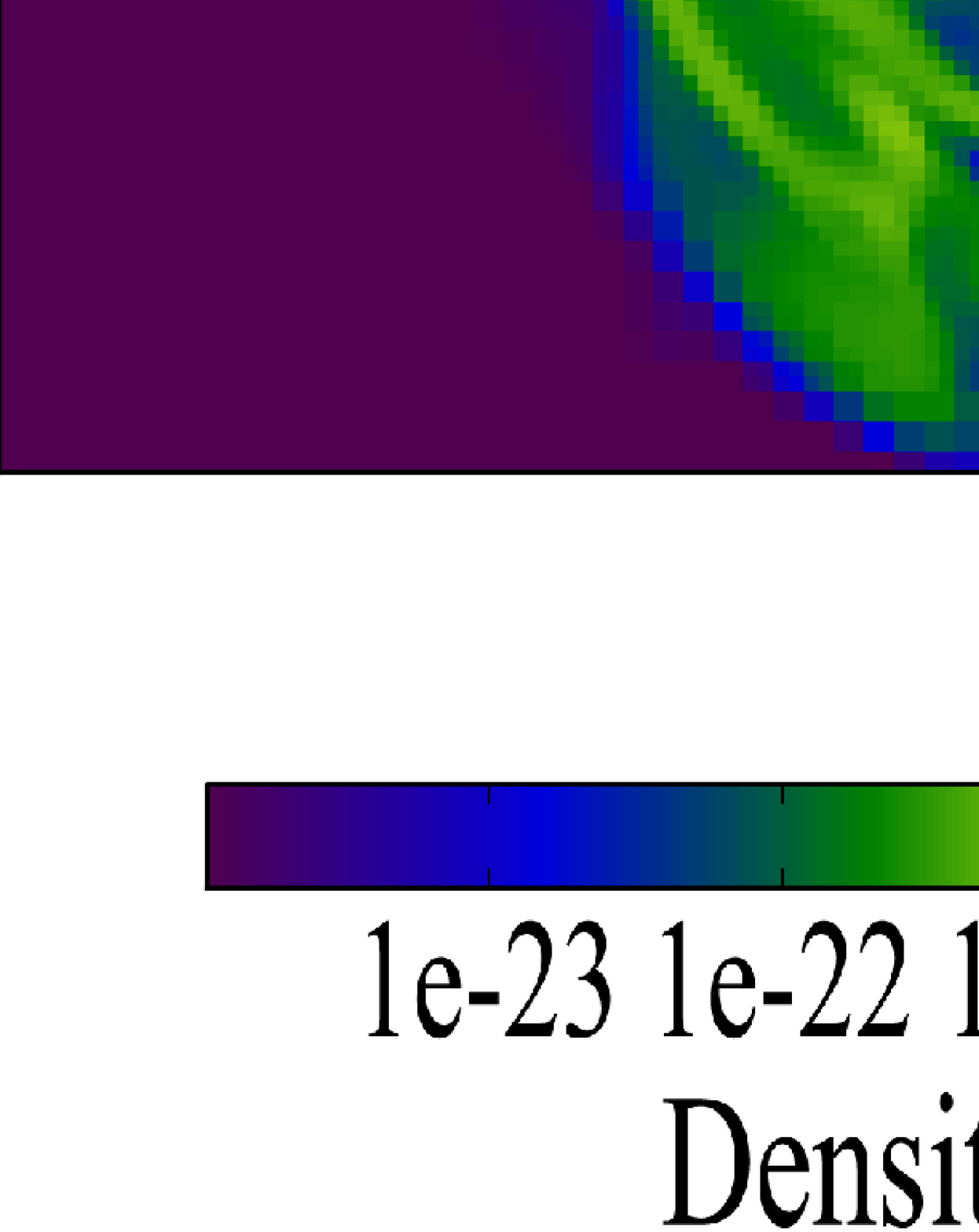}
\vspace{0.05cm}
\includegraphics[angle=0,width=8.3cm]{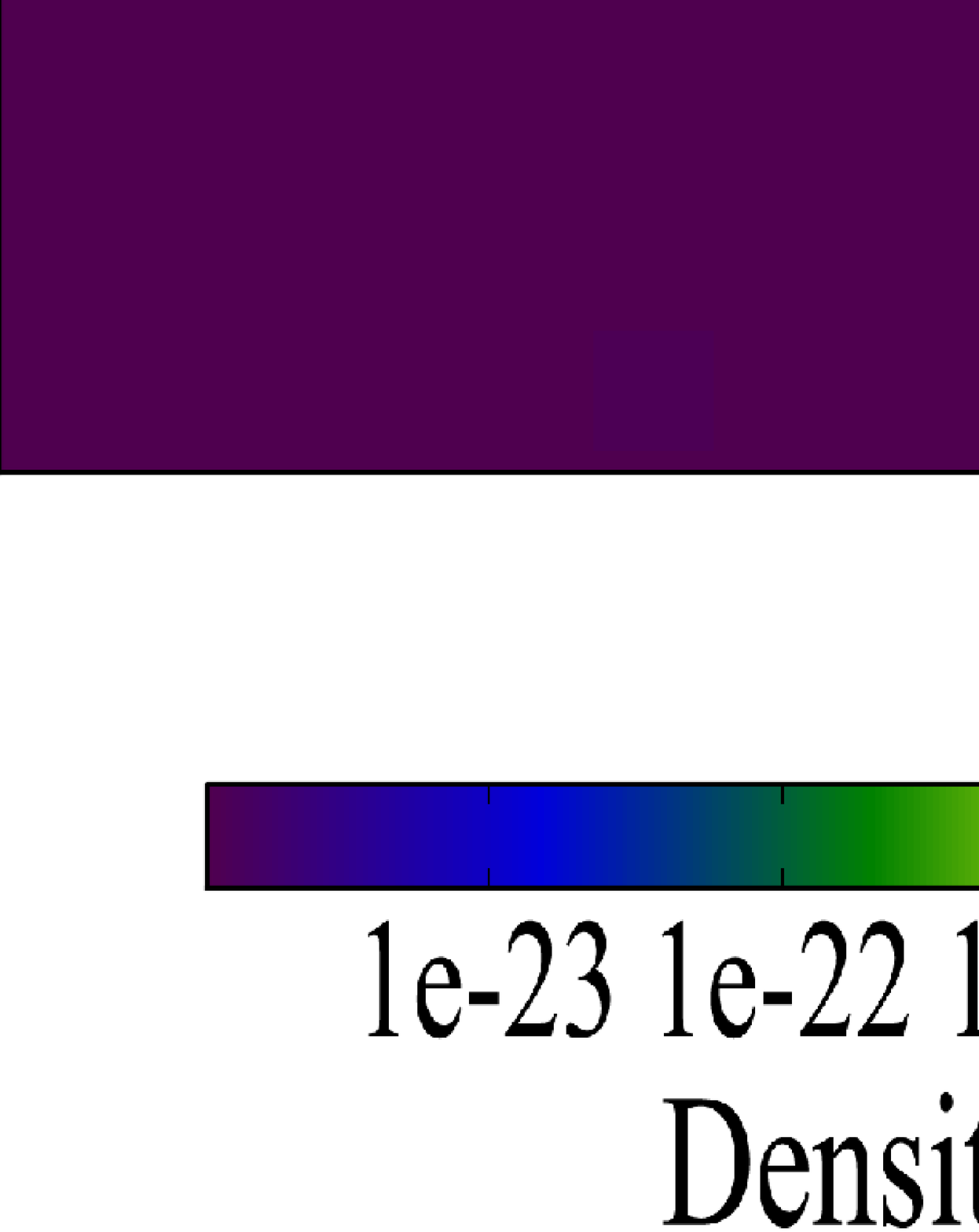}
\caption{From left to right: Slices of density, temperature and Jeans mass of the central 1 kpc of a halo at redshift 5 for metallicities of  Z$/$Z$_{\odot}$ = 10$^{-4}$ (top),   Z$/$Z$_{\odot}$ = 10$^{-2}$ (middle), and  Z$/$Z$_{\odot}$ = 10$^{-1}$ (bottom) in the x-z (left column) and x-y (right column) plane for a radiation field of G$_0$ = 10$^{-2}$.} \label{fig3}
\end{figure*}

\begin{figure}
\includegraphics[angle=0,width=8cm]{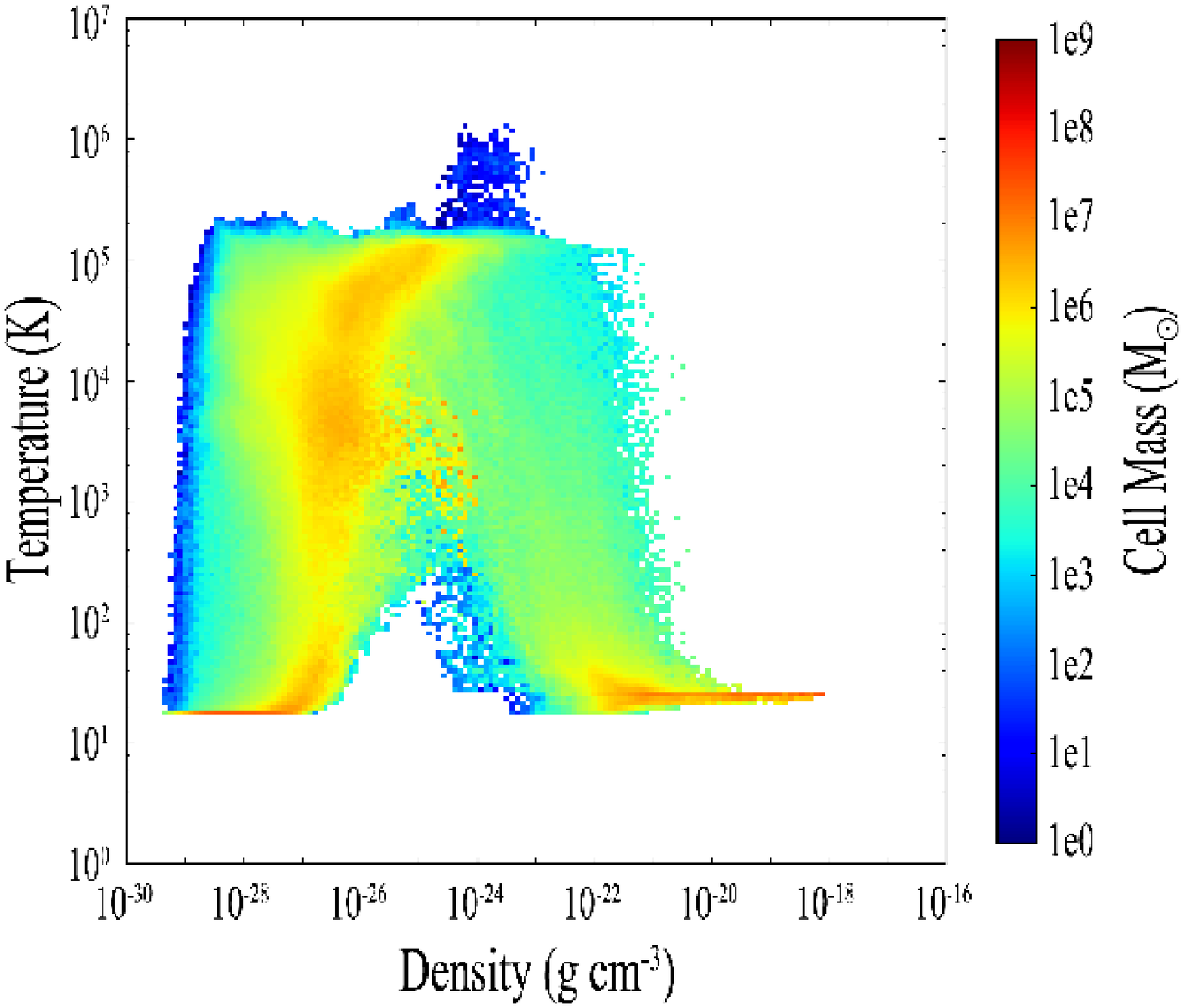}
\caption{Density-temperature profile of the central 50 kpc of a halo at redshift z$\sim$5 for a metallicity of Z$/$$\rm{Z_{\odot}}$=10$^{-1}$ and a radiation field of $\rm{G_0}$=10$^{-2}$.} \label{fig4}
\end{figure}

In our cosmological simulations for the Z2$-$G1 and Z4$-$G1 runs  we find (as expected) that gas in the high-metallicity case cools down to lower temperatures and has higher densities than in the low-metallicity case. This is shown in Figure \ref{fig1} where we plot the density$-$temperature profile of a halo at redshift $z\sim$5 for metallicities of Z$/$$Z_{\odot}$ = 10$^{-4}$ (left) and $Z/Z_{\odot}$=10$^{-2}$ (right) and a radiation field of $G_0$=10$^{-2}$. In the high-metallicity case there is more gas in the central 50 kpc. This is because the cooling is more efficient, so that the gas has less pressure support against gravity and the halo accretes more material. When we look at the high densities ($\rho$ $>$10$^{-23}$ g cm$^{-3}$) in these plots it is clear that in the high-metallicity case there is a lot more gas at low temperatures than in the low-metallicity case. This gas is identified as the cold and dense gas phase of the multi-phase ISM. The key feature here is that the redshift at which a multi-phase ISM is established depends on metallicity. This is shown in Figure \ref{fig2} where we plot the density$-$temperature profile of the same halo as shown in Figure \ref{fig1} at $z\sim$20 for metallicities of  $Z/Z_{\odot}$ = 10$^{-4}$ (left) and $Z/Z_{\odot}$ = 10$^{-2}$ (right). We see here that the cold dense phase is completely lacking in the Z4$-$G1 run whereas it is already present in the Z2$-$G1 run. This means that metal-rich gas cools more efficiently and therefore the multi-phase ISM is established at earlier times in the higher metallicity case than in the lower metallicity case. These results are consistent with those of \cite{2009ApJ...696.1065J}, although they did not include background radiation.\\ 
\indent Due to the shorter cooling time a halo in the metal-rich case evolves dynamically faster and becomes more compact at $z =$ 5 compared to a metal-poorer halo. The effect of metallicity on a halo can be seen in Figure \ref{fig3}, which shows slices of the density, temperature, and the Jeans mass ($M_J$) of the central 1 kpc of a halo for the Z4$-$G1 (top), Z2$-$G1 (middle), and Z1$-$G1 (bottom) runs in the $x-z$ (left column) and $x-y$ (right column) planes at redshift 5. We compute the Jeans mass as
\begin{eqnarray}
M\rm{_{J}} = (\frac{5 k_{b}}{T}{G\mu m_H})^{3/2}(\frac{3}{4\pi\rho})^{1/2},
\end{eqnarray}
where $\rho$ is the mass density, $G$ is the gravitational constant, $k_{b}$ is the Boltzmann constant, $T$ is the temperature, $\mu$ is the mass per particle, and $m_H$ is the hydrogen mass.
Because the high-metallicity gas ($Z/Z_{\odot}$ $\geq$ 10$^{-2}$) cools to lower temperatures, the Jeans mass in those runs is two orders of magnitude smaller than in the low-metallicity ($Z/Z_{\odot}$ = 10$^{-4}$) case. We also see that the Jeans masses for metallicities $Z/Z_{\odot}$ = 10$^{-2}$ and $Z/Z_{\odot}$ = 10$^{-1}$ are comparable. This is because above a  metallicity of about $Z/Z_{\odot}$ = 10$^{-2}$ the cooling efficiency of ambient gas no longer increases strongly with a rise in metallicity \citep{2008ApJ...678L...5S}. In Figure \ref{fig4}, we plot the density$-$temperature profile of the central 50 kpc of a halo at $z \sim$ 5 for the Z1$-$G1 run. The phase diagrams of the runs Z1$-$G1 (Figure \ref{fig4}) and Z2$-$G1 (Figure \ref{fig1}, right) are indeed similar.\\
\subsection{UV Background Radiation} 
\begin{figure*}
\includegraphics[angle=0,width=8cm]{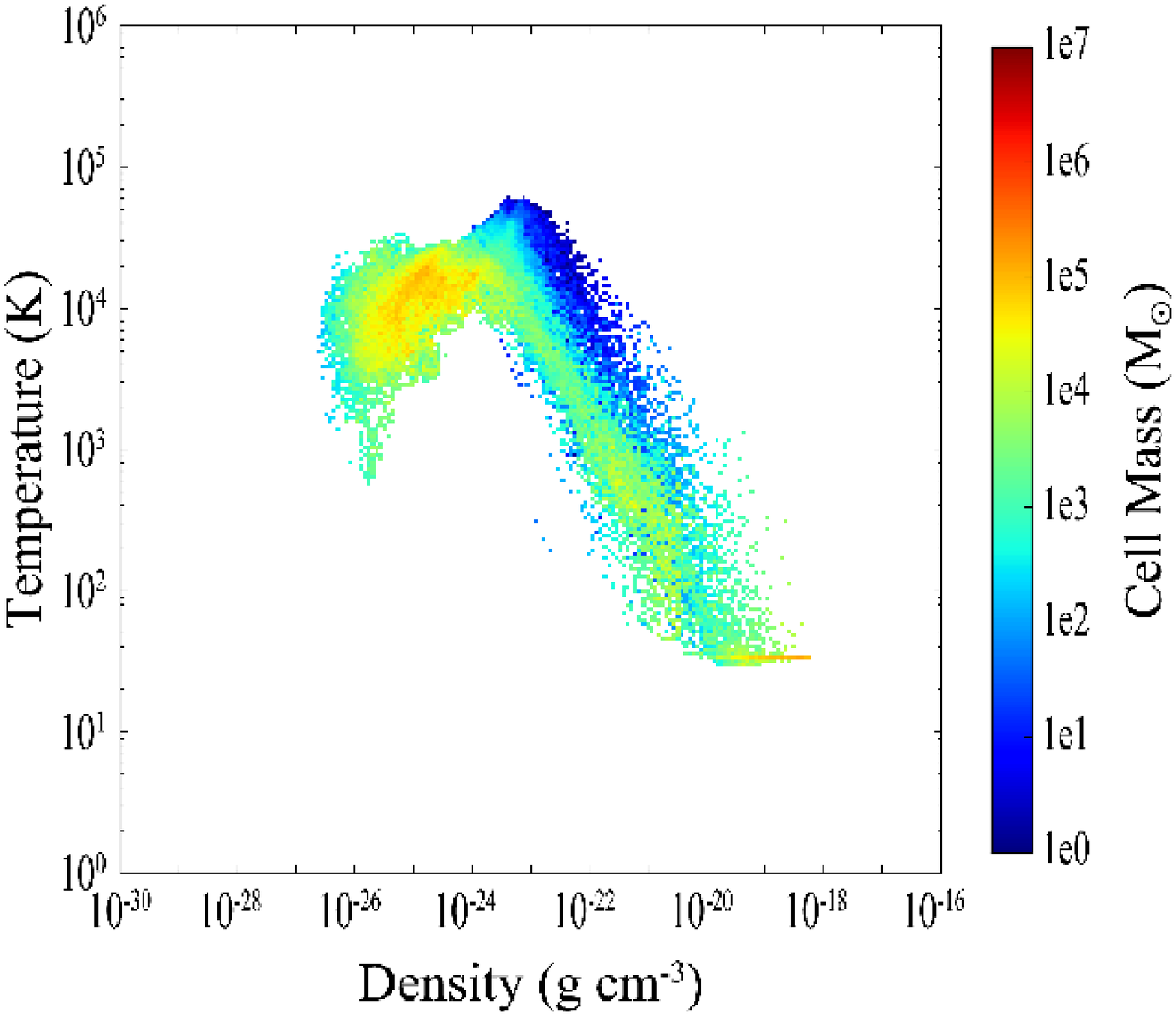}
\hspace{0.5cm}
\includegraphics[angle=0,width=8cm]{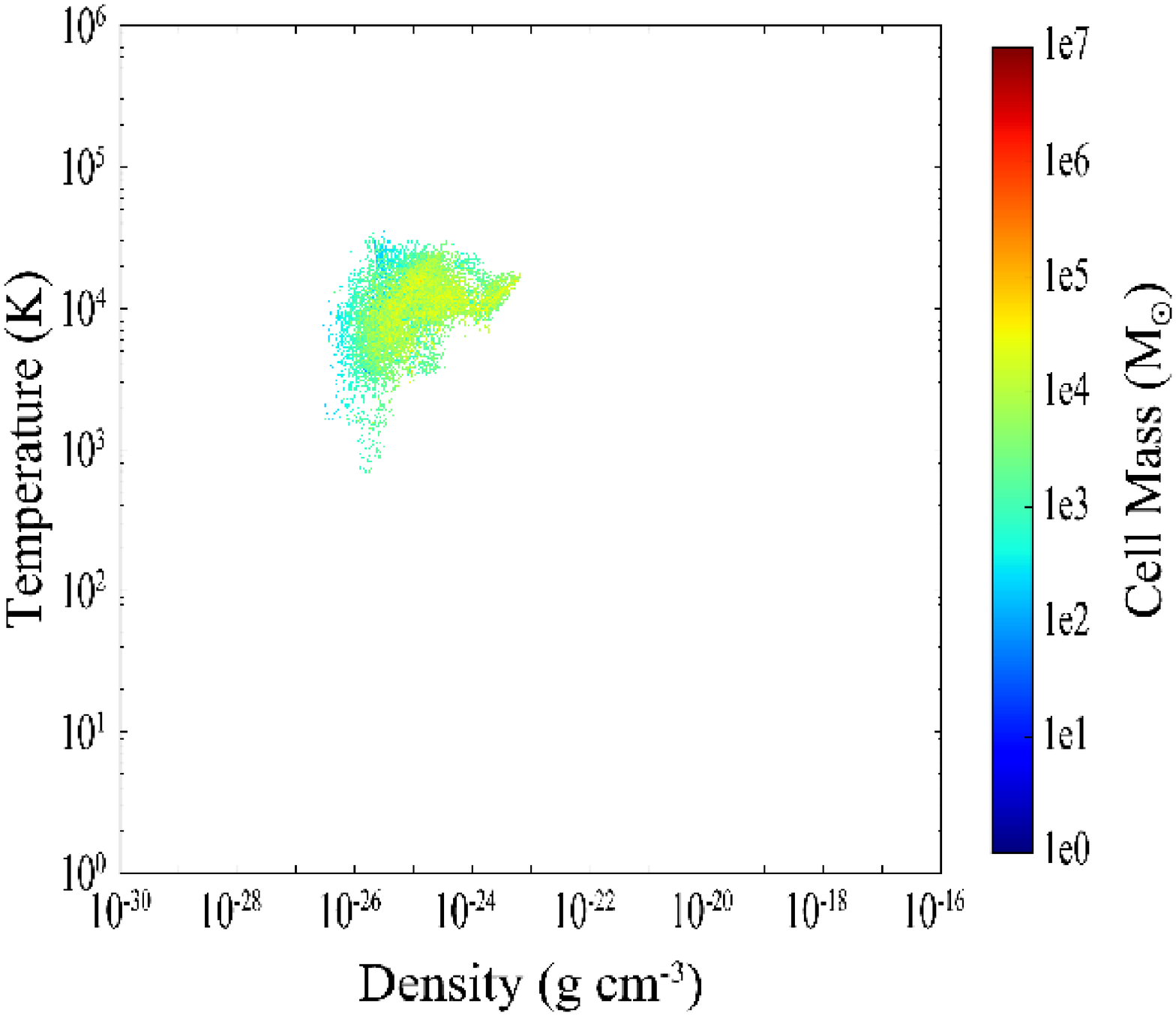}\\
\newline
\includegraphics[angle=0,width=8cm]{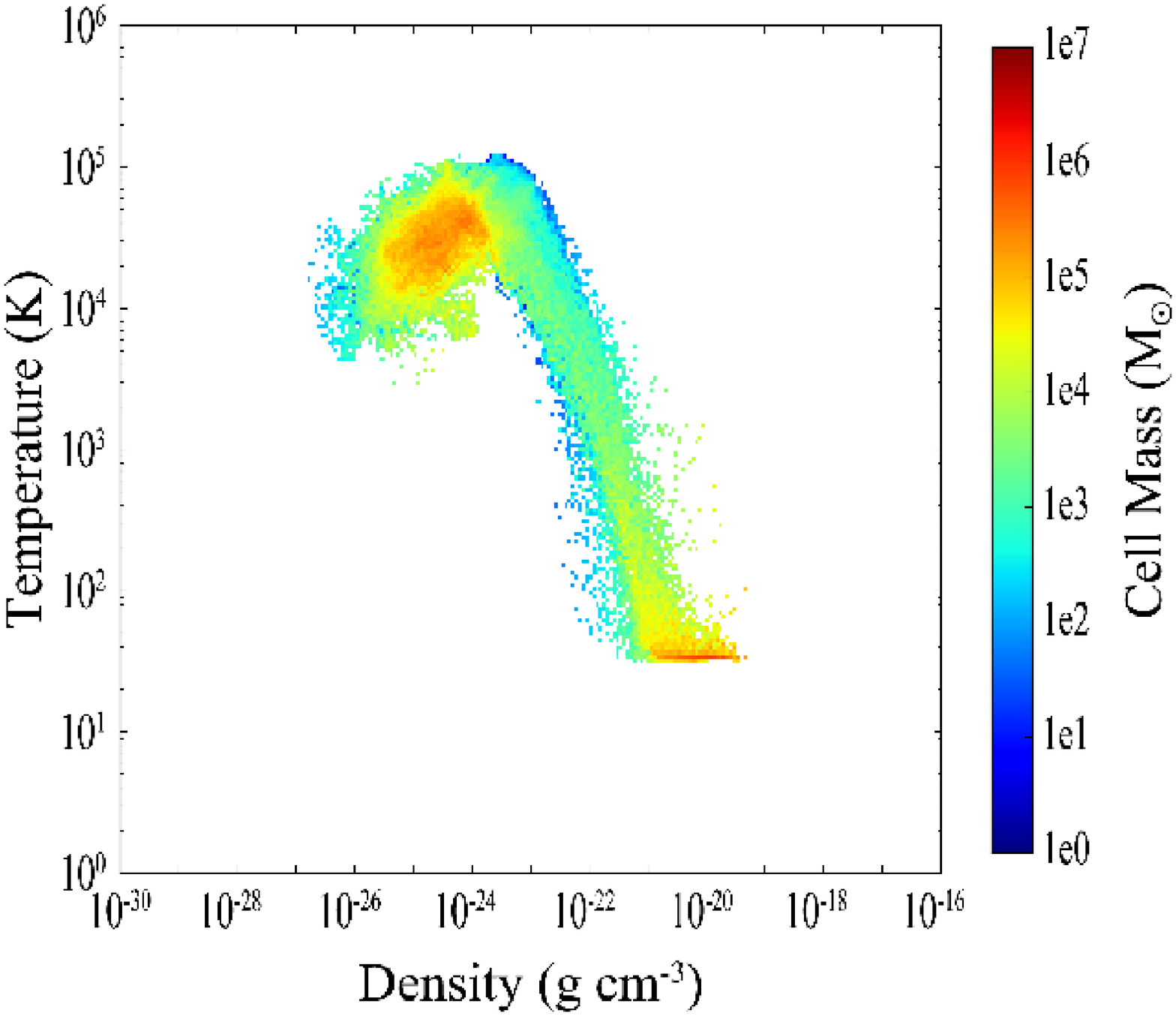}
\hspace{0.5cm}
\includegraphics[angle=0,width=8cm]{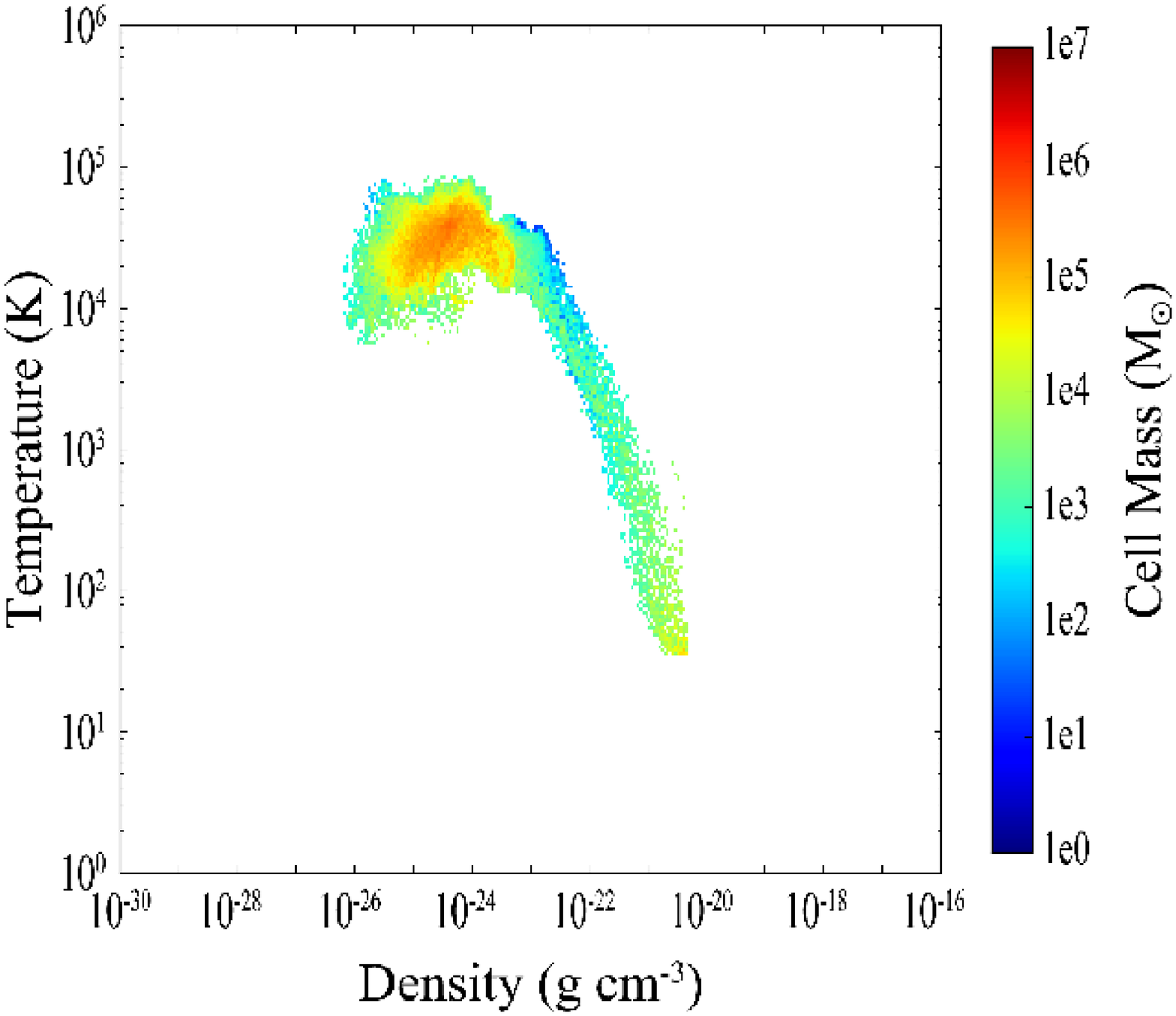}
\caption{Density$-$temperature profile of the central 2 kpc of a halo at redshift 7.88 for a metallicity of $Z/Z_{\odot}$ = 10$^{-3}$ with radiation fields of  $G_0$ = 10$^{-2}$ (top left) and $G_0$ = 10$^{-1}$ (top right), and a metallicity of $Z/Z_{\odot}$ = 10$^{-2}$ with radiation fields of $G_0$ = 10$^{-2}$ (bottom left) and $G_0$ = 10$^{-1}$ (bottom right).}\label{fig5}
\end{figure*}

\begin{figure*}
\centering
\includegraphics[angle=0,width=6cm]{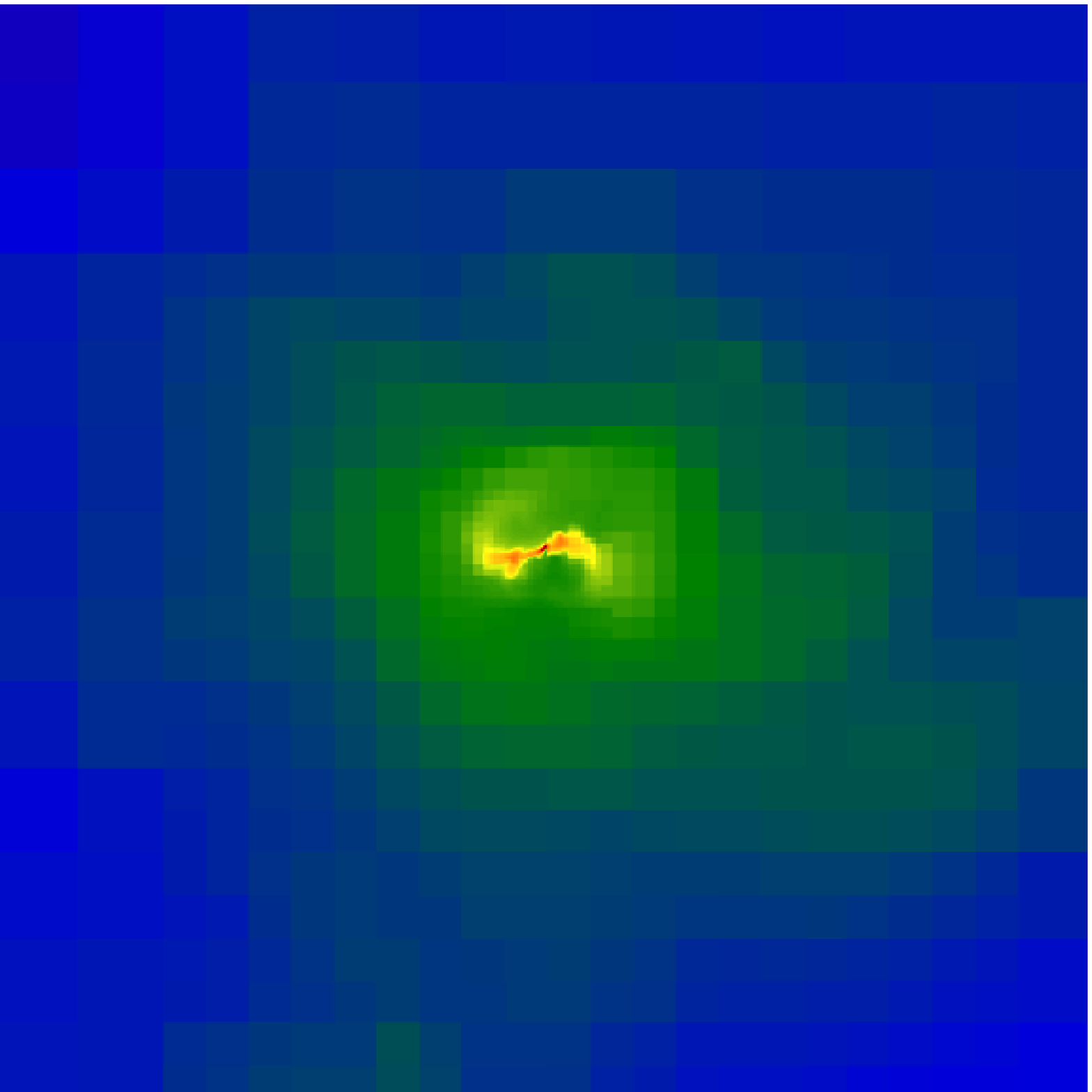}
\vspace{0.05cm}
\includegraphics[angle=0,width=7.57cm]{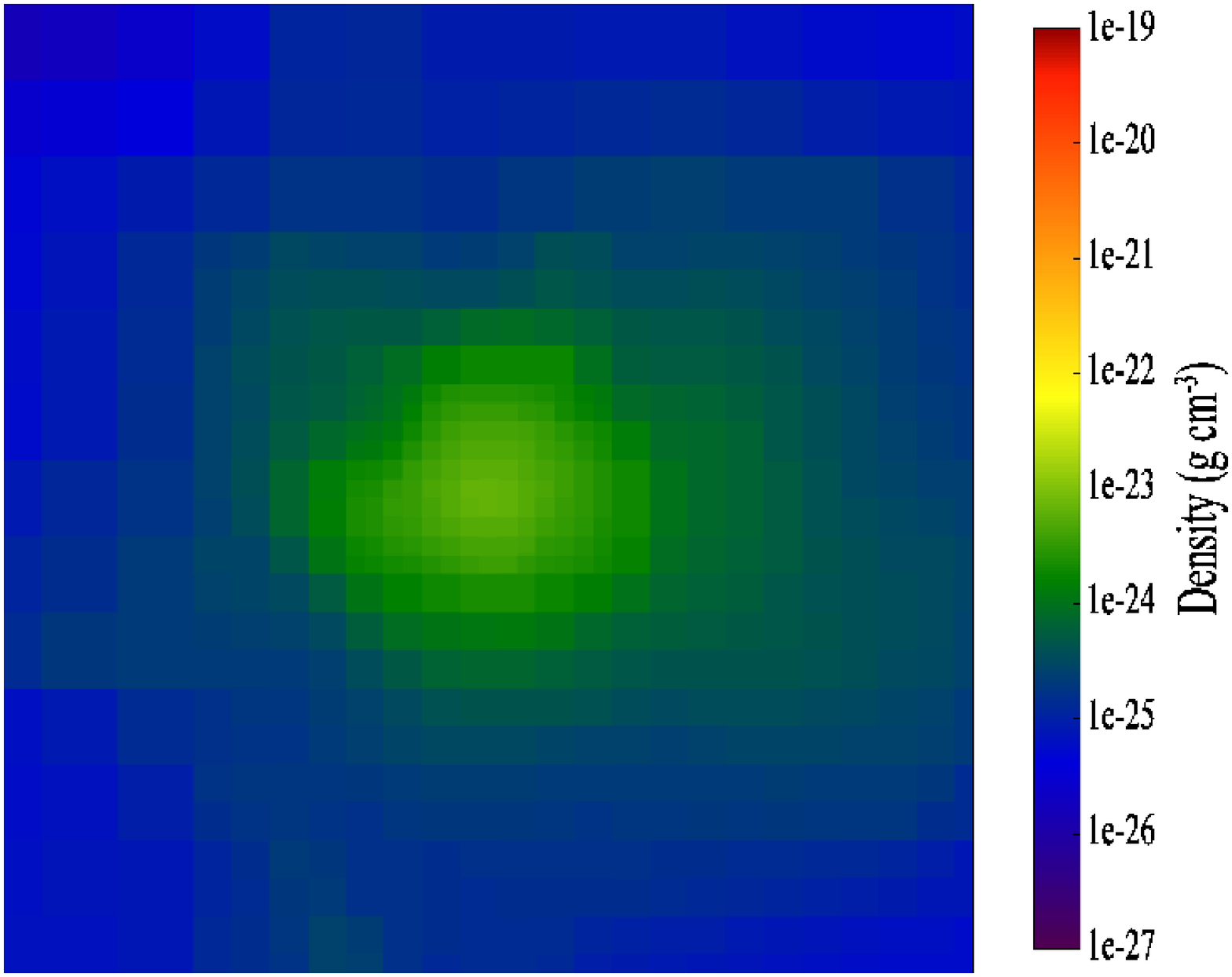}\\
\includegraphics[angle=0,width=6.cm]{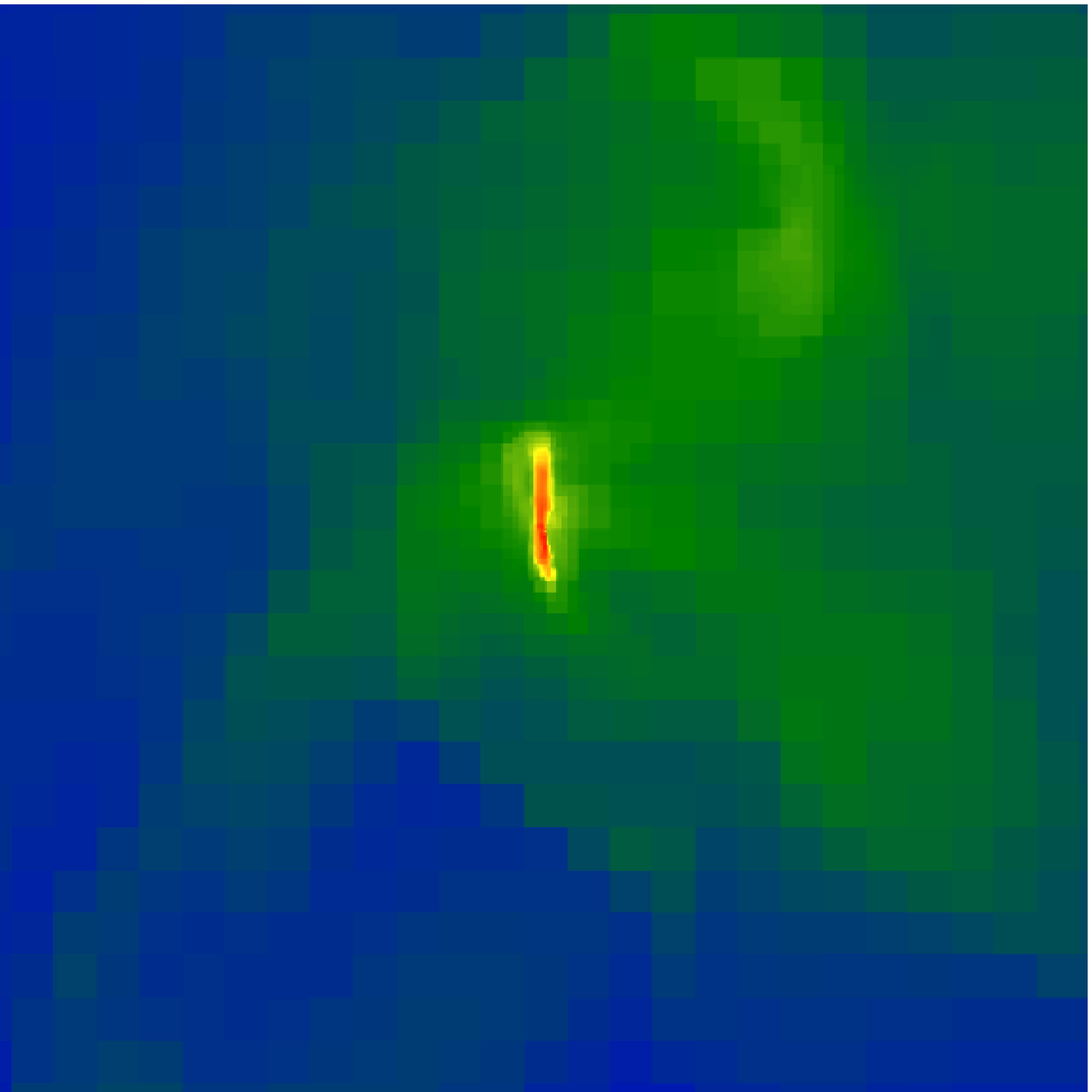}
\vspace{0.05cm}
\includegraphics[angle=0,width=7.57cm]{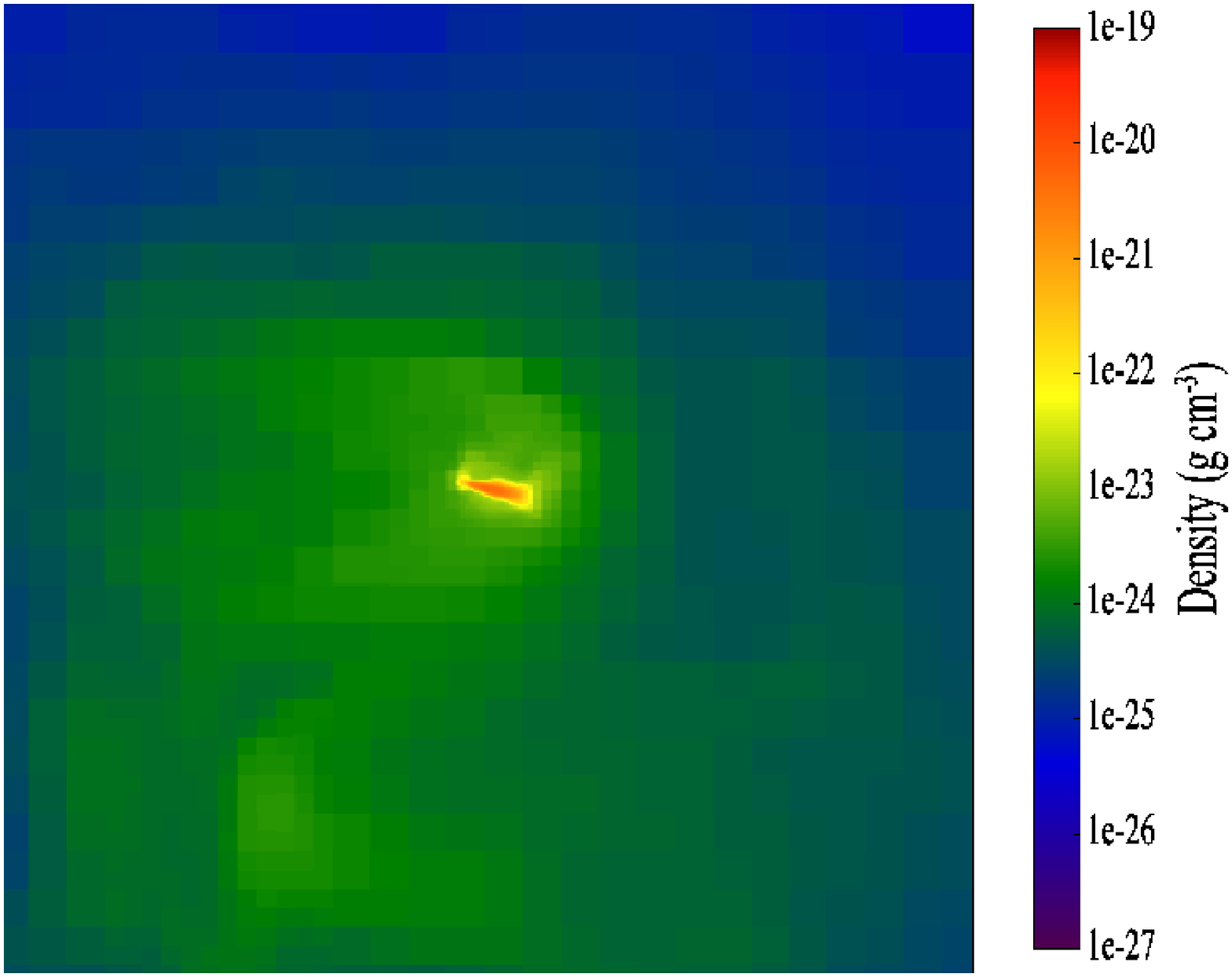}
\caption{Density slices of the central 2 kpc of a halo at redshift $z =$ 7.88 for a metallicity of $Z/Z_{\odot} = $10$^{-3}$ with radiation fields of $G_0$ = 10$^{-2}$ (top left) and $G_0$ = 10$^{-1}$ (top right), and a metallicity of  $Z/Z_{\odot}$ = 10$^{-2}$ with radiation fields of $G_0$ = 10$^{-2}$ (bottom left) and $G_0$ = 10$^{-1}$ (bottom right) in the $x-z$ plane.} \label{fig6}
\end{figure*}

\begin{figure}
\includegraphics[angle=0,width=8cm]{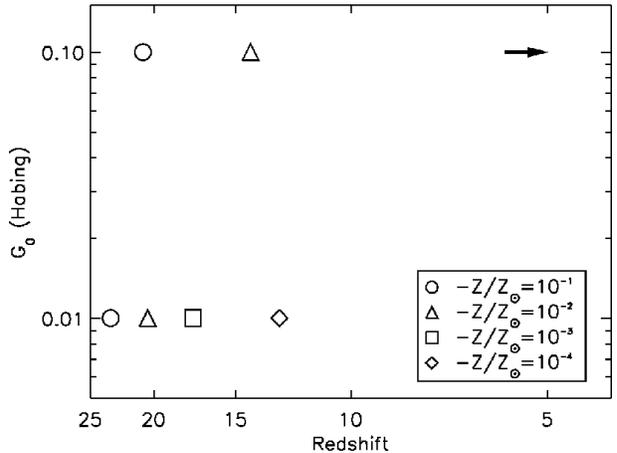}
\caption {Radiation field strength vs. redshift. Symbols mark the redshifts where a cold dense gas phase is established. The arrow represents the runs with $Z/Z_\odot$=10$^{-3}$ and $Z/Z_\odot$=10$^{-4}$, for a radiation field of $G_0$=10$^{-1}$, which do not develop a cold dense gas phase before $z =$ 5.}\label{fig7}
\end{figure}

In order to see the effect of a UV radiation field on the evolution of gas we performed multiple simulations with different background radiation fields for the same metallicity. In Figure \ref{fig5}, we plot the density$-$temperature profile of the central 2 kpc of a halo at $z \sim$ 8 for metallicities of $Z/Z_{\odot}$ = 10$^{-3}$ (top) and $Z/Z_{\odot}$ = 10$^{-2}$ (bottom) with two different background radiation fields of $G_0$ = 10$^{-2}$ (left) and $G_0$ = 10$^{-1}$ (right). When we increase the radiation field for a metallicity of $Z/Z_{\odot}$ = 10$^{-3}$, the cold and dense phase does not form anymore, which is clearly seen in the phase diagrams of Figure \ref{fig5} : At densities above $\rho$ = 10$^{-23}$ g cm$^{-3}$ there is no cold gas present anymore in the Z3-G10 run (top right). This is because gas cannot cool down as efficiently as in the lower radiation field case due to more heat input and dissociation of H$_2$ and CO. Thus, the halo takes more time to collapse and become condensed. In our simulations we followed the halo until redshift $z =$ 5, and in the Z3$-$G10 run the gas has still not collapsed at this redshift. On the other hand, in the $Z/Z_{\odot}$ = 10$^{-2}$ case the cold dense phase survives regardless of irradiation. This indicates that the cold dense phase is fragile to UV radiation for metallicities of $Z/Z_{\odot}$ $\leq$ 10$^{-3}$ and robust to UV radiation for metallicities of $Z/Z_{\odot}$ $\geq$ 10$^{-2}$.\\ 
\indent As is shown in Figure  \ref{fig6}, where we plot the density profile of the central 2 kpc of a halo for the Z3$-$G1 (top left), Z3$-$G10 (top right), Z2$-$G1 (bottom left), and Z2$-$G10 (bottom right) runs, the effect of UV irradiation on the model galaxy morphology is large for $Z/Z_\odot$ = 10$^{-3}$ and quite modest for $Z/Z_\odot$ = 10$^{-2}$. In fact, we see that in the case of low metallicity and strong radiation field (Z3$-$G10, top right) the compact disk-like structure is no longer formed, while in the higher metallicity and strong radiation field (Z2$-$G10, bottom right) the disk-like structure is still present. This is because the high-metallicity gas is able to cool faster so that it collapses and reaches higher densities earlier in its evolution, and thus builds up column density to protect the cold and dense phase.\\
\indent In Figure \ref{fig7}, we plot the redshift at which a multi-phase ISM is established depending on the metallicity and the radiation field. We find that, to first order, the cold and dense phase survives if the ratio $F_0$/$Z$ is smaller than $\sim$10$^{-2}$ erg cm$^{-2}$ s$^{-1}$ in solar units.\\
\indent All this has a significant influence on the subsequent star formation. The Jeans mass sets a typical value for a cloud to collapse gravitationally and fragment, depending on both temperature and density of the medium. Although the metal-enriched gas has a higher cooling efficiency than metal-poor gas, under the influence of a background radiation field the gas will stay hot and the Jeans mass will remain high. Therefore, the UV radiation background is a key parameter to take into account  when studying the transition from Pop III to Pop II stars. In our simulations we see that including a constant radiation background raises the critical density value for this transition from $Z_{cr}$ $\sim$ 10$^{-3.5}$$Z_{\odot}$ to $Z_{cr}$ $\sim$ 10$^{-2}$$Z_{\odot}$ if $F_0$$>$10$^{-2}$ erg s$^{-1}$ cm$^{-2}$. \\
\subsection{Dynamics} 
\begin{figure*}
\includegraphics[angle=0,width=5.0cm]{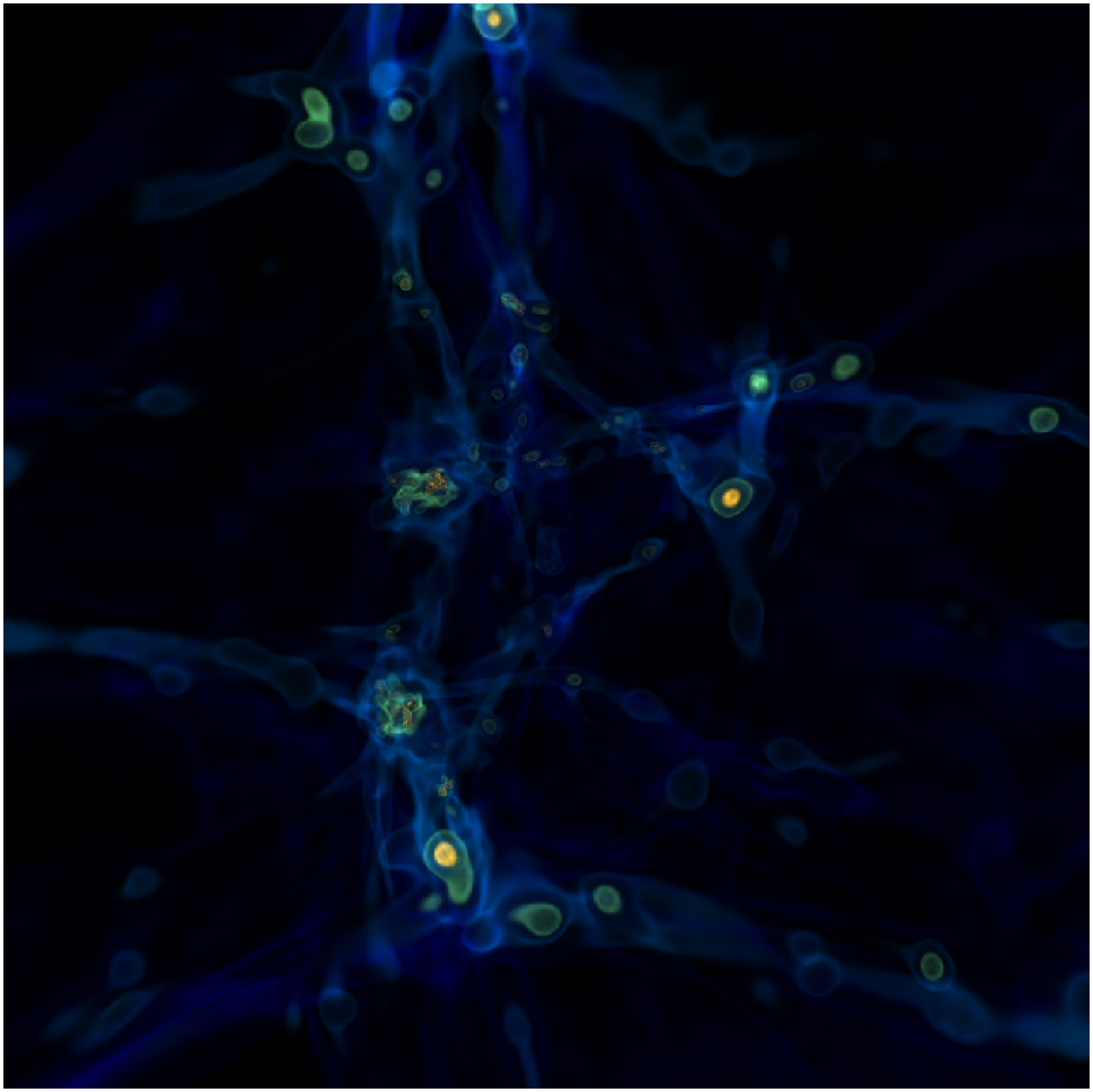}
\vspace{0.05cm}
\includegraphics[angle=0,width=5.0cm]{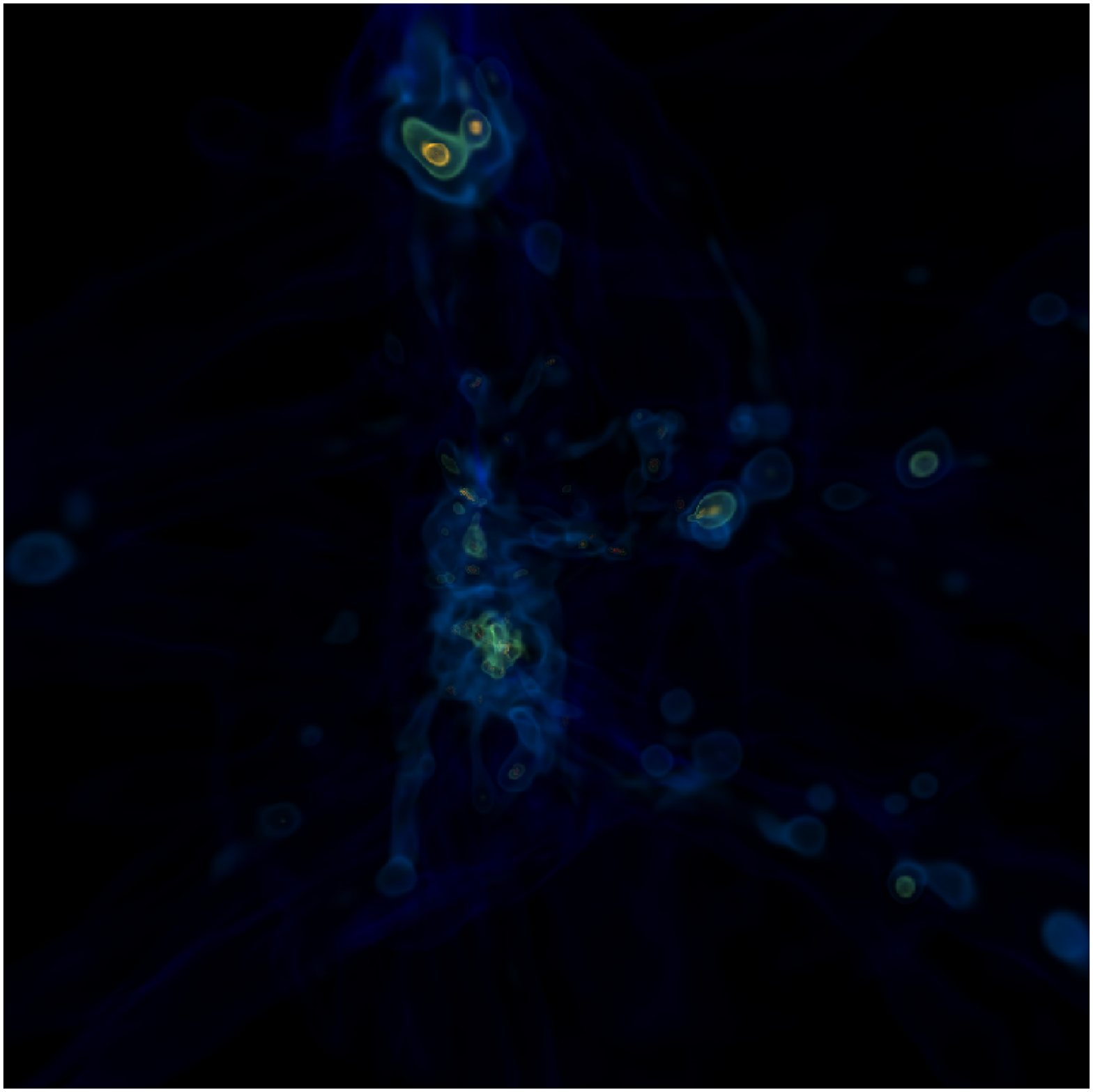}
\vspace{0.05cm}
\includegraphics[angle=0,width=6.22cm]{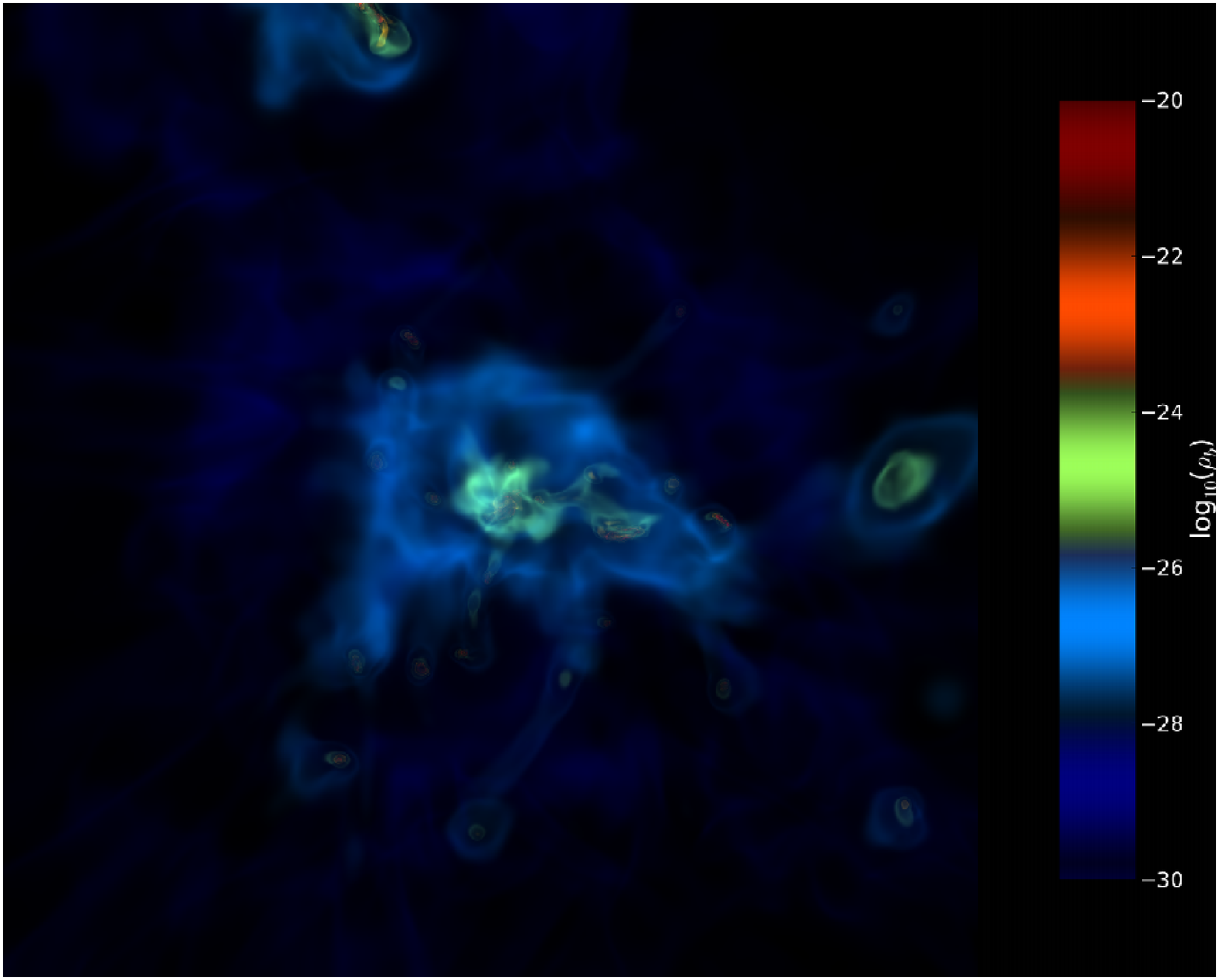}\\
\includegraphics[angle=0,width=5.0cm]{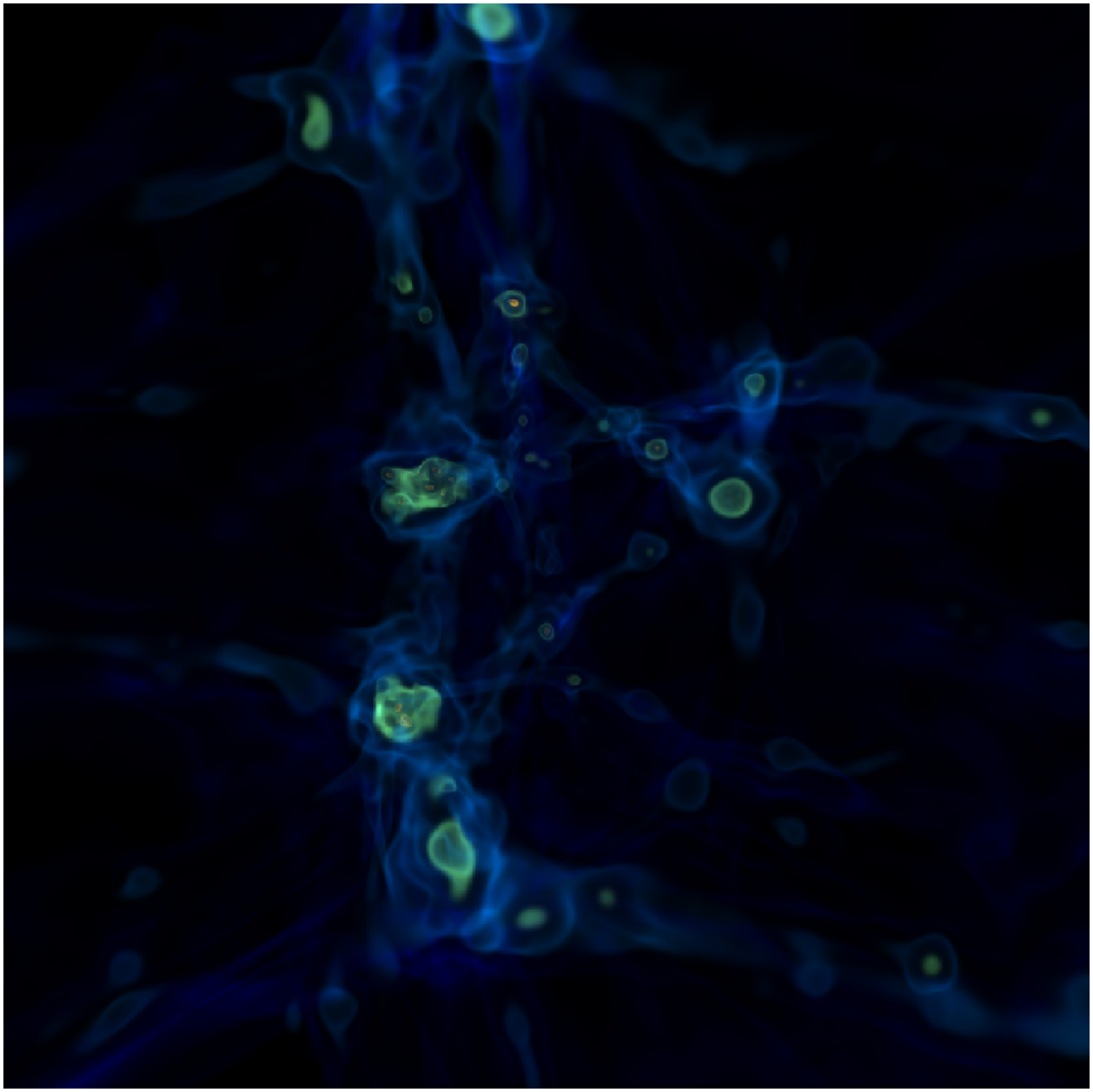}
\vspace{0.05cm}
\includegraphics[angle=0,width=5.0cm]{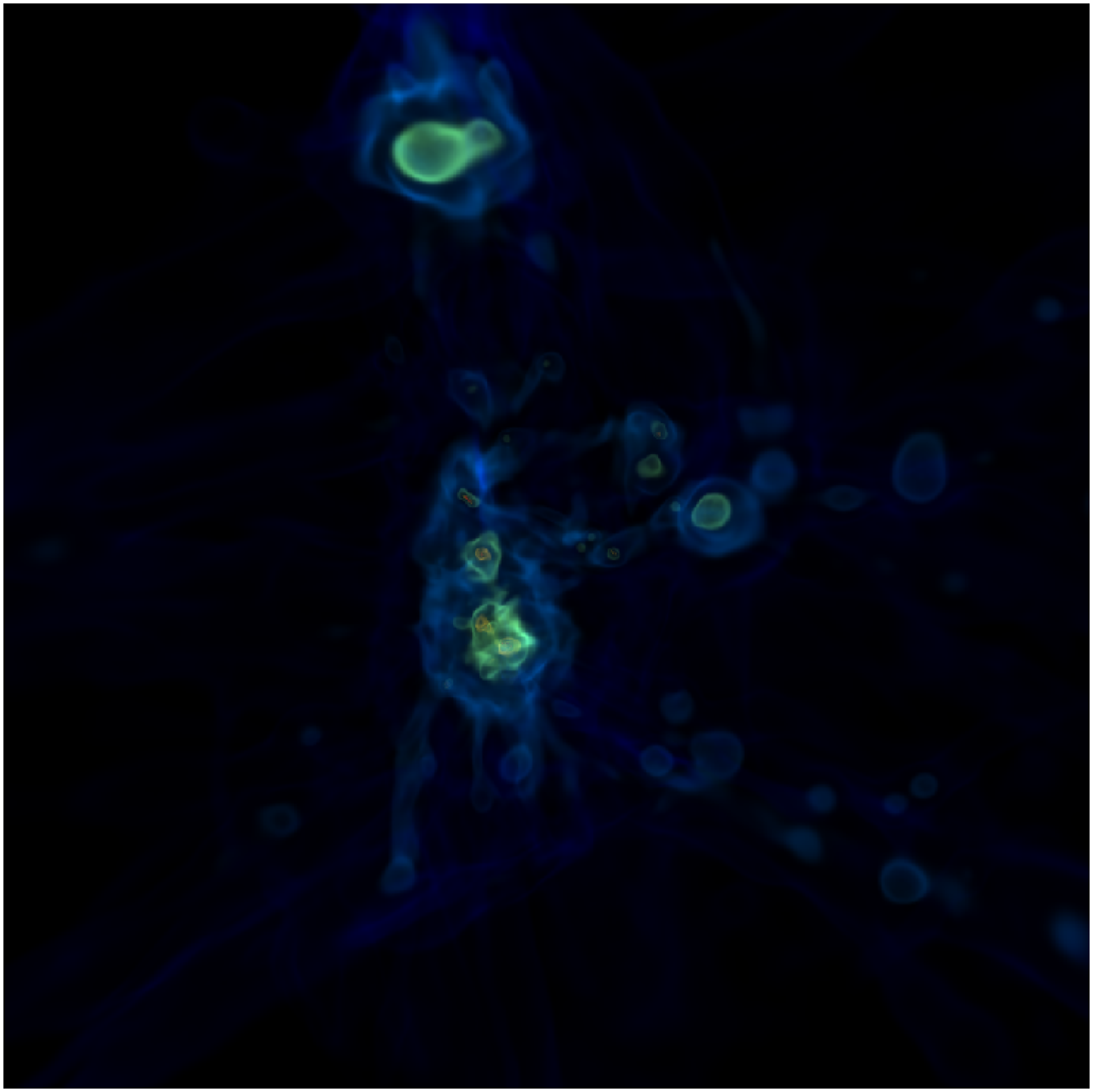}
\vspace{0.05cm}
\includegraphics[angle=0,width=6.22cm]{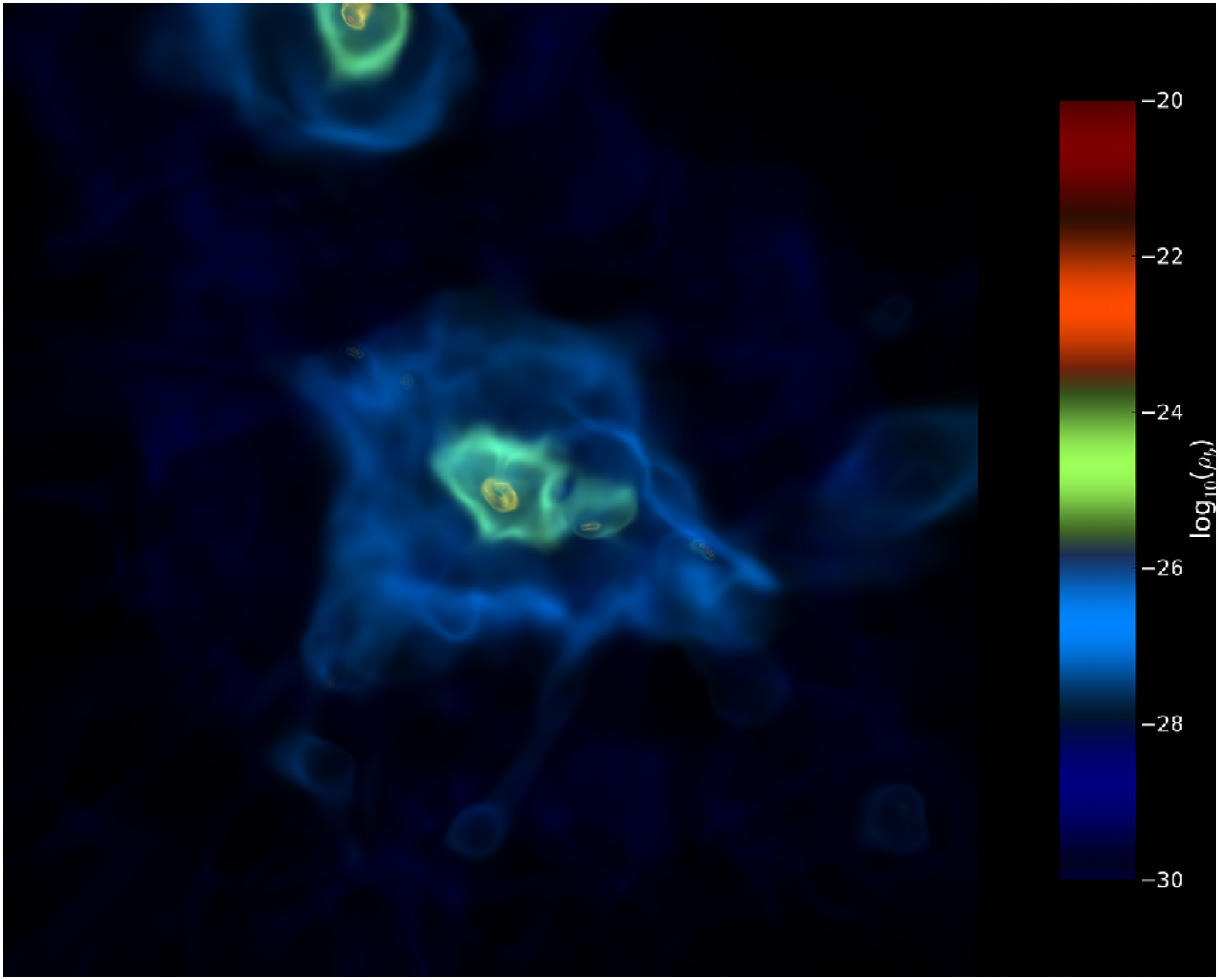}\\
\includegraphics[angle=0,width=5.0cm]{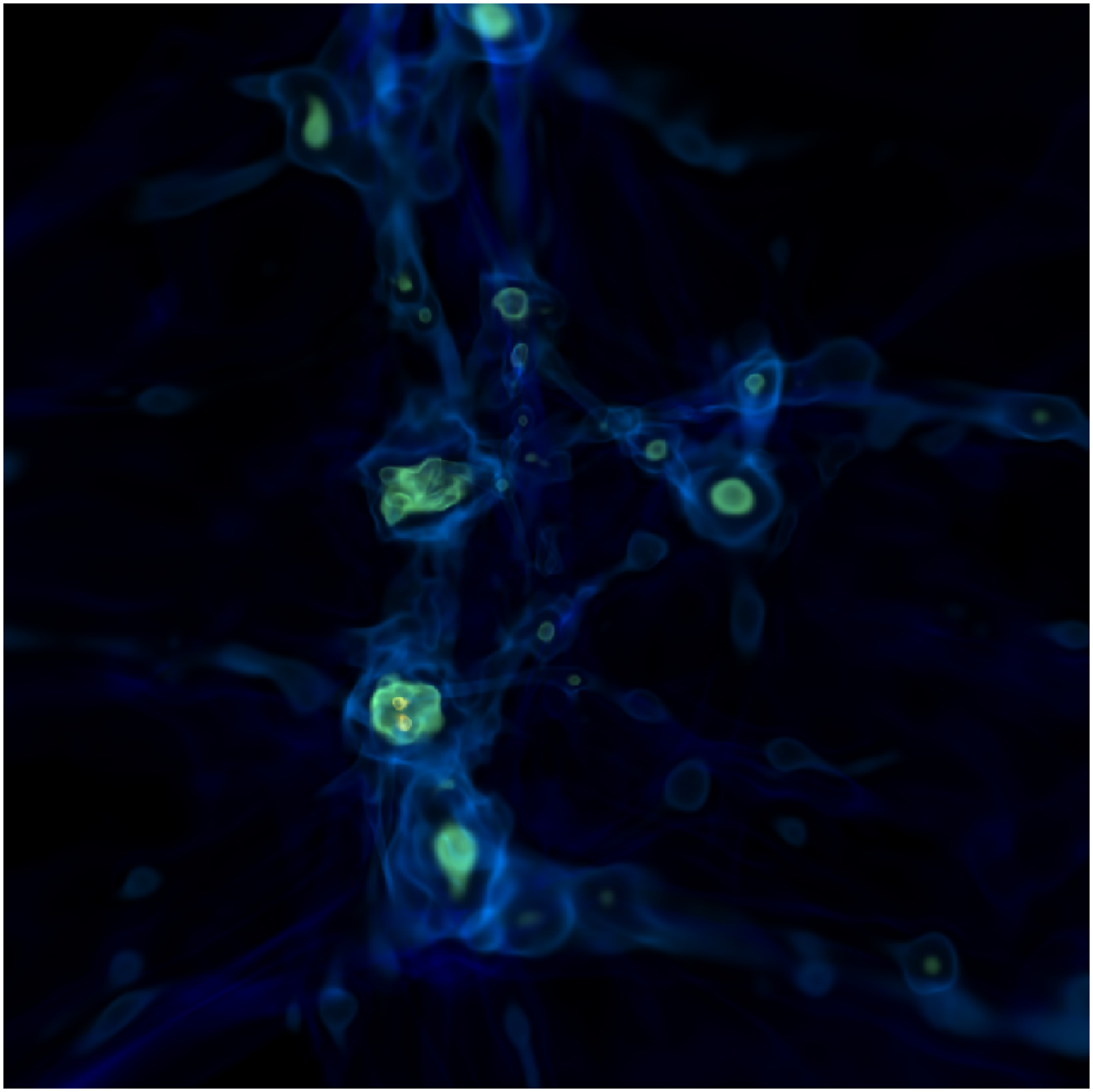}
\vspace{0.05cm}
\includegraphics[angle=0,width=5.0cm]{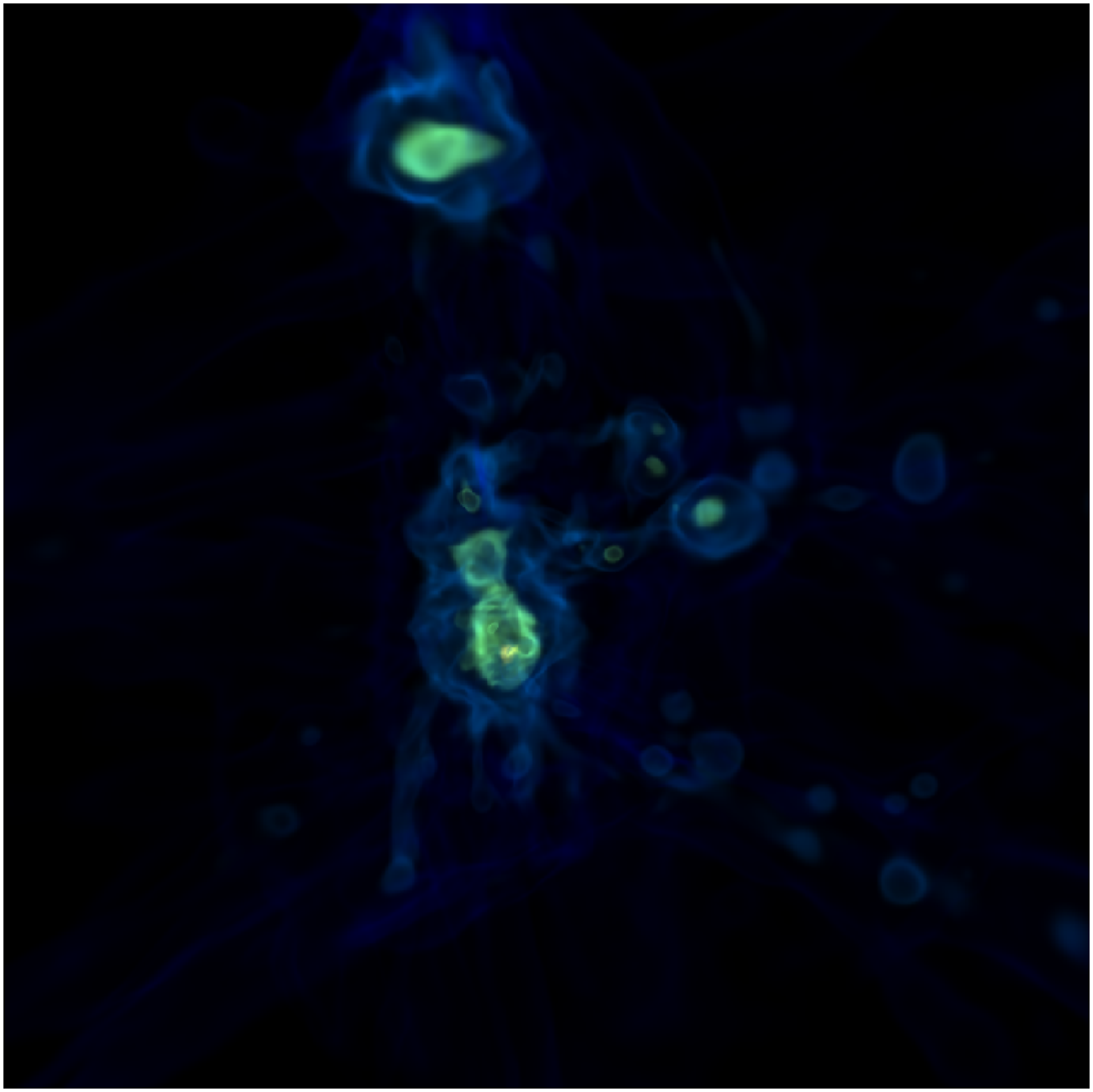}
\vspace{0.05cm}
\includegraphics[angle=0,width=6.22cm]{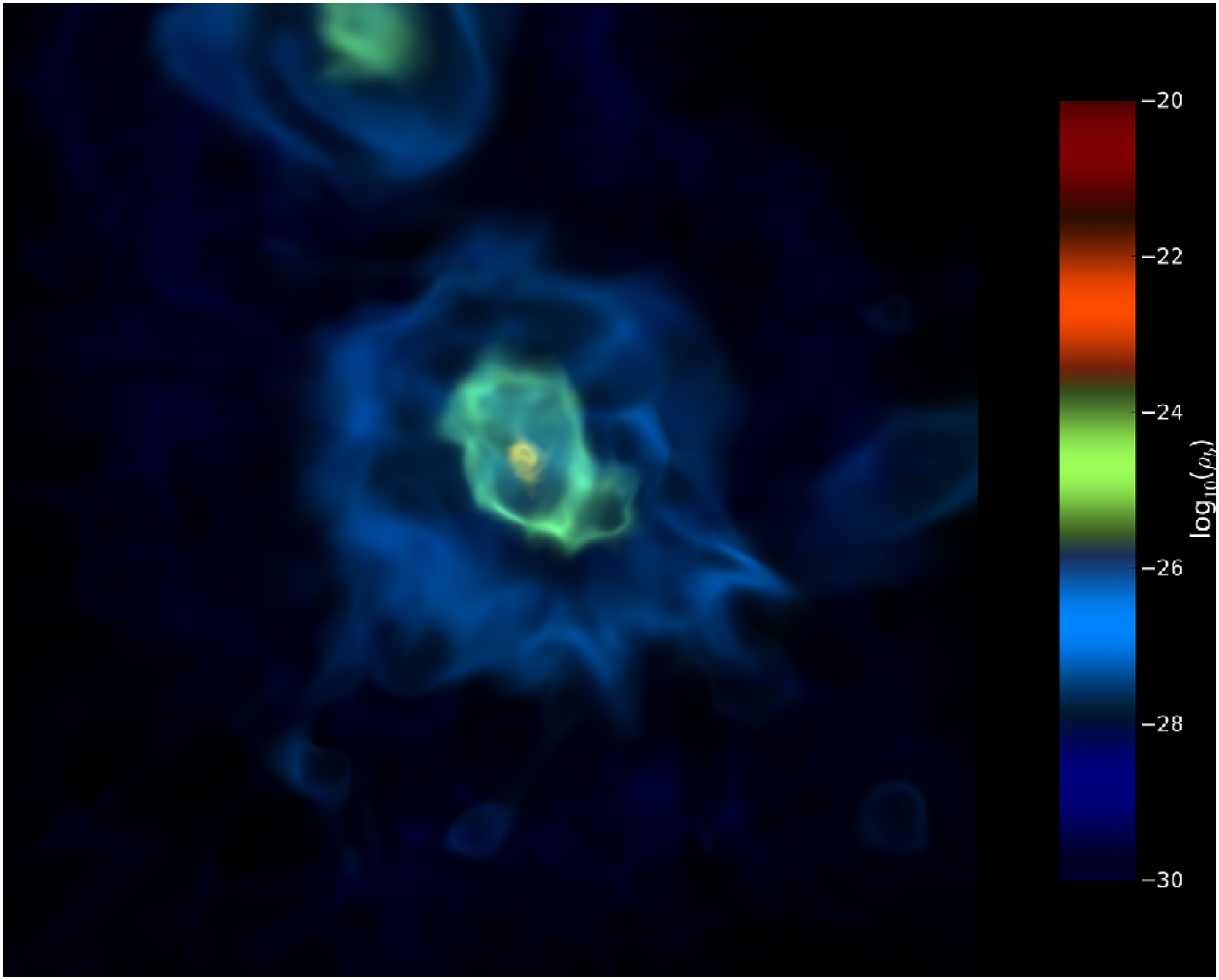}
\caption{Three-dimensional, volume rendered, density snapshots of the runs for metallicities of $Z/Z_{\odot}$ = 10$^{-1}$ (top) and $Z/Z_{\odot}$ = 10$^{-2}$ (middle) for a radiation field of $G_0$ = 10$^{-2}$ and $Z/Z_{\odot}$ = 10$^{-2}$ (bottom) for a radiation field of $G_0$ = 10$^{-1}$ at redshifts $z =$ 7.88 (left), $z =$ 6.24 (middle), and $z =$ 5 (right). Size of the box for redshifts 7.88 and 6.24 is 60 kpc, and for redshift 5 is 10 kpc. }\label{fig8}
\end{figure*}

\begin{figure}
\includegraphics[angle=0,width=8cm]{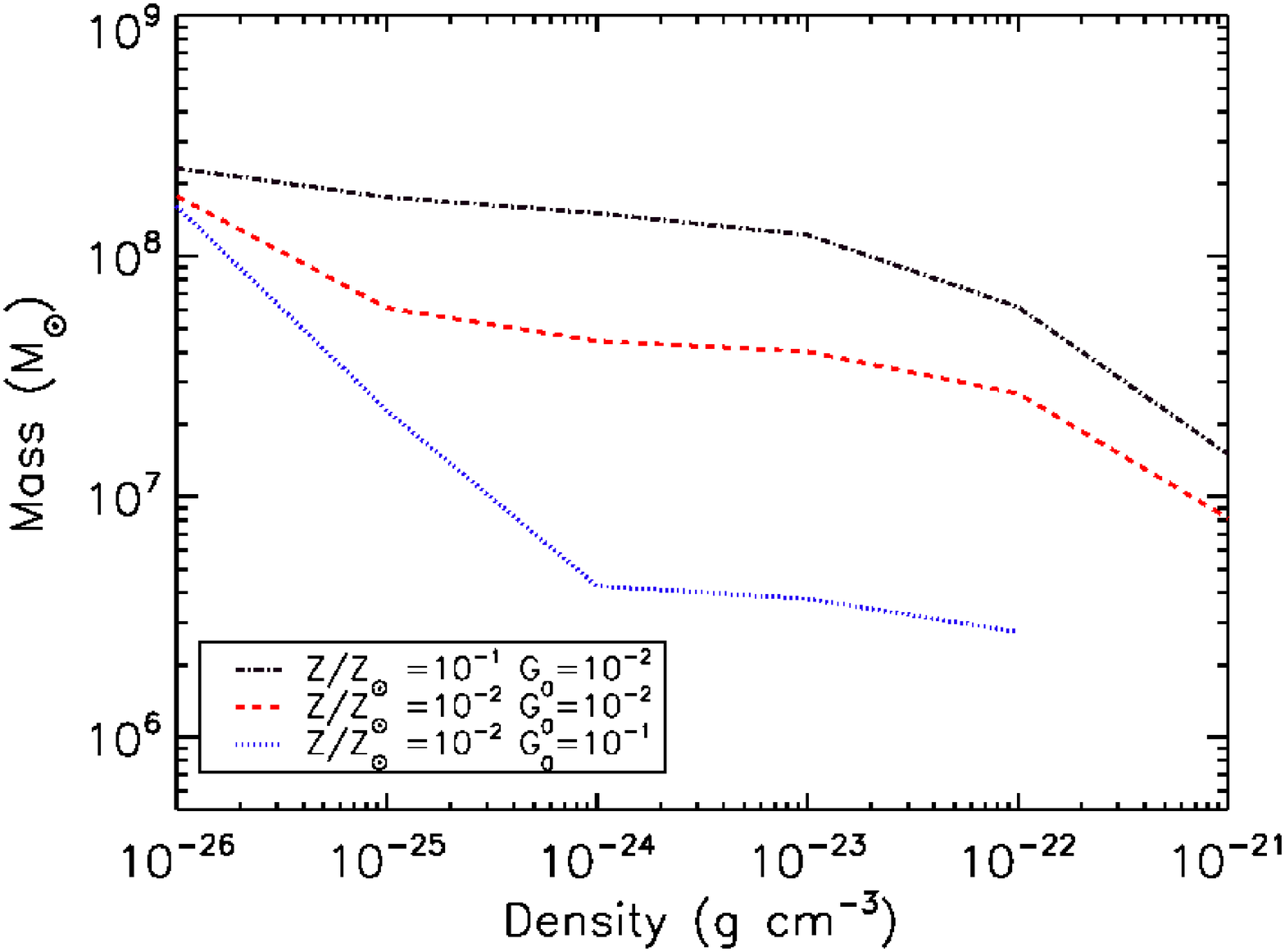}
\includegraphics[angle=0,width=8cm]{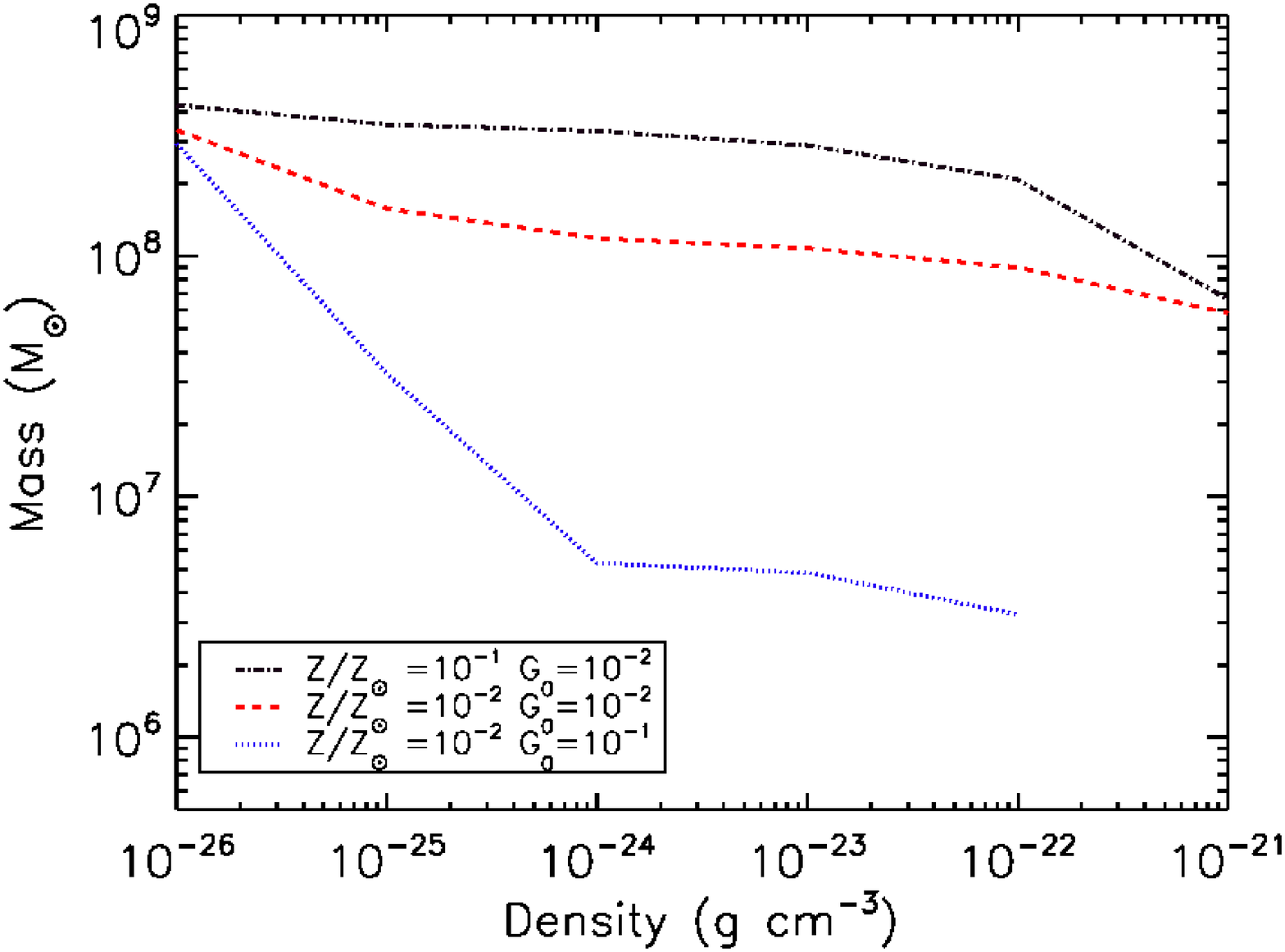}
\includegraphics[angle=0,width=8cm]{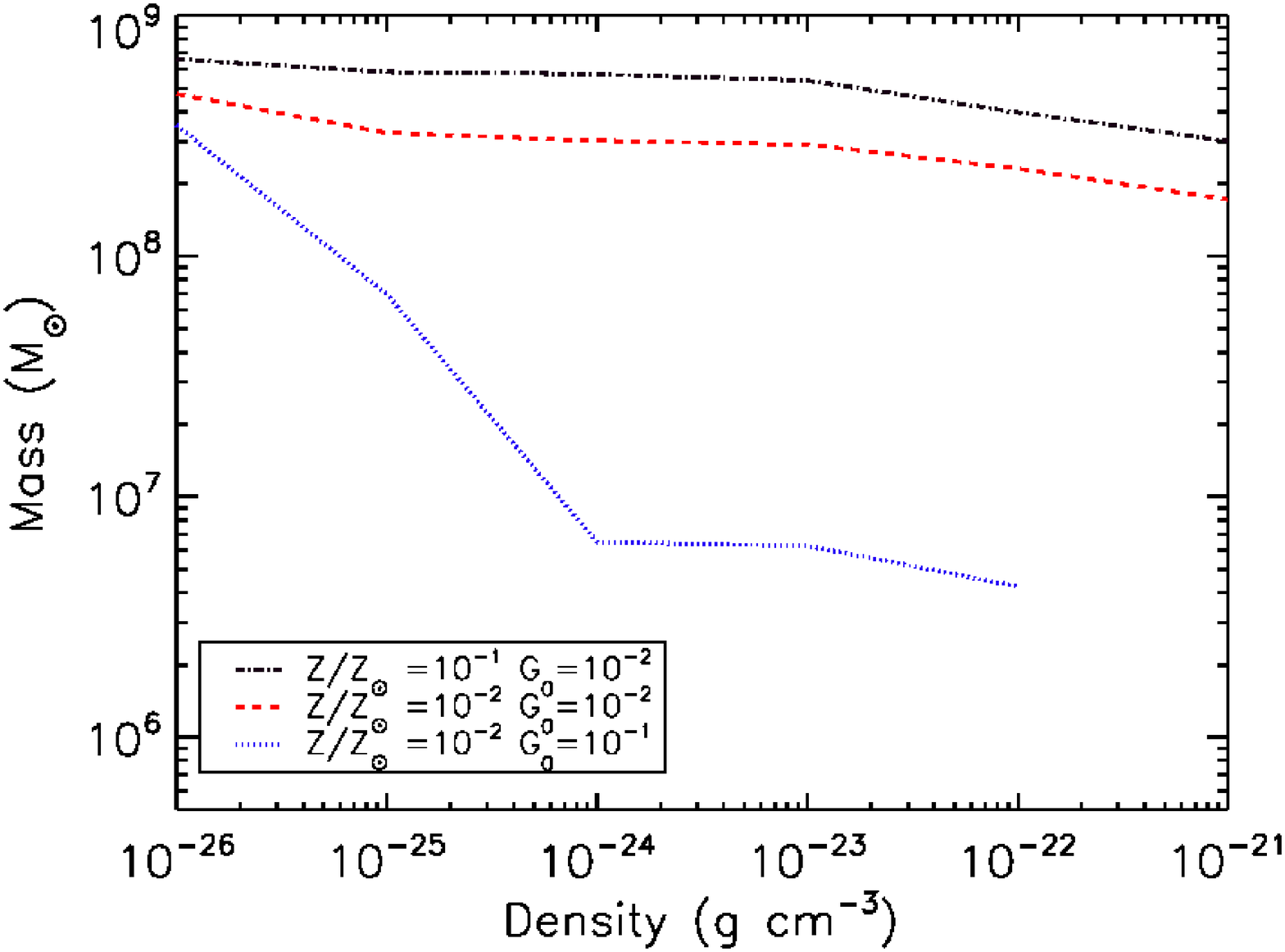}
\caption{Gas mass vs. density threshold of the central 10 kpc for metallicities of $Z/Z_{\odot}$ = 10$^{-1}$ and $Z/Z_{\odot}$ = 10$^{-2}$ for a radiation field of $G_0$=10$^{-2}$ and $Z/Z_{\odot}$ = 10$^{-2}$ for a radiation field of G$_0$ = 10$^{-1}$ at redshifts $z =$ 7.88 (top), $z =$ 6.24 (middle), and $z =$ 5 (bottom).}\label{fig9}
\end{figure}

\begin{figure}
\includegraphics[angle=0,width=8cm]{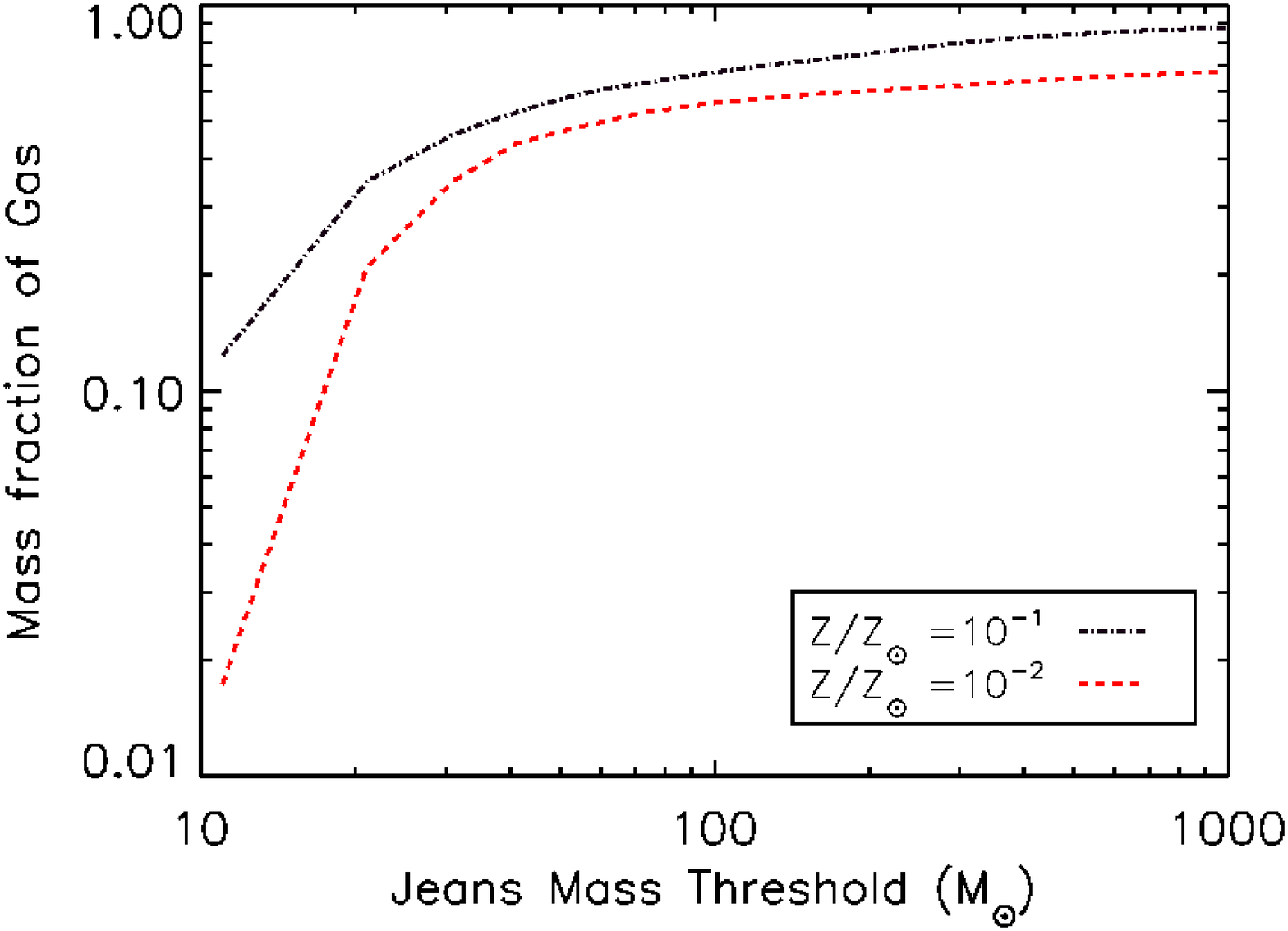}
\caption{Gas mass fraction vs. Jeans mass threshold for metallicities of $Z/Z_{\odot}$ = 10$^{-1}$ (black dots) and $Z/Z_{\odot}$ = 10$^{-2}$ (red dashes) at redshift $z =$ 5 for the central 1 kpc.}\label{fig10}
\end{figure}

\begin{figure}
\includegraphics[angle=0,width=8cm]{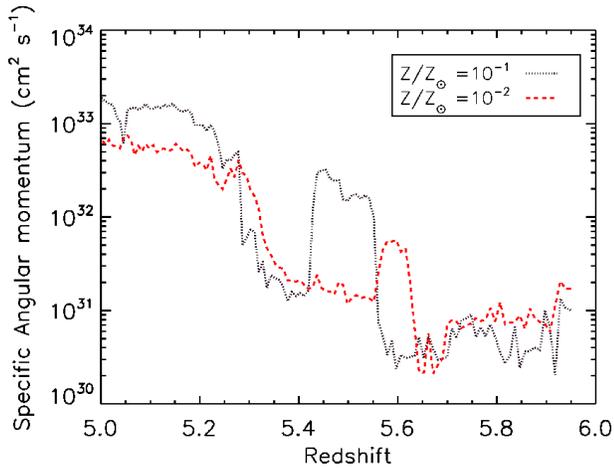}
\caption{Specific angular momentum vs. redshift for metallicities  of $Z/Z_{\odot}$ = 10$^{-1}$ (black dots) and $Z/Z_{\odot}$ = 10$^{-2}$ (red dashes) for the central 1 kpc.}\label{fig11}
\end{figure}

\begin{figure}
\includegraphics[angle=0,width=7cm]{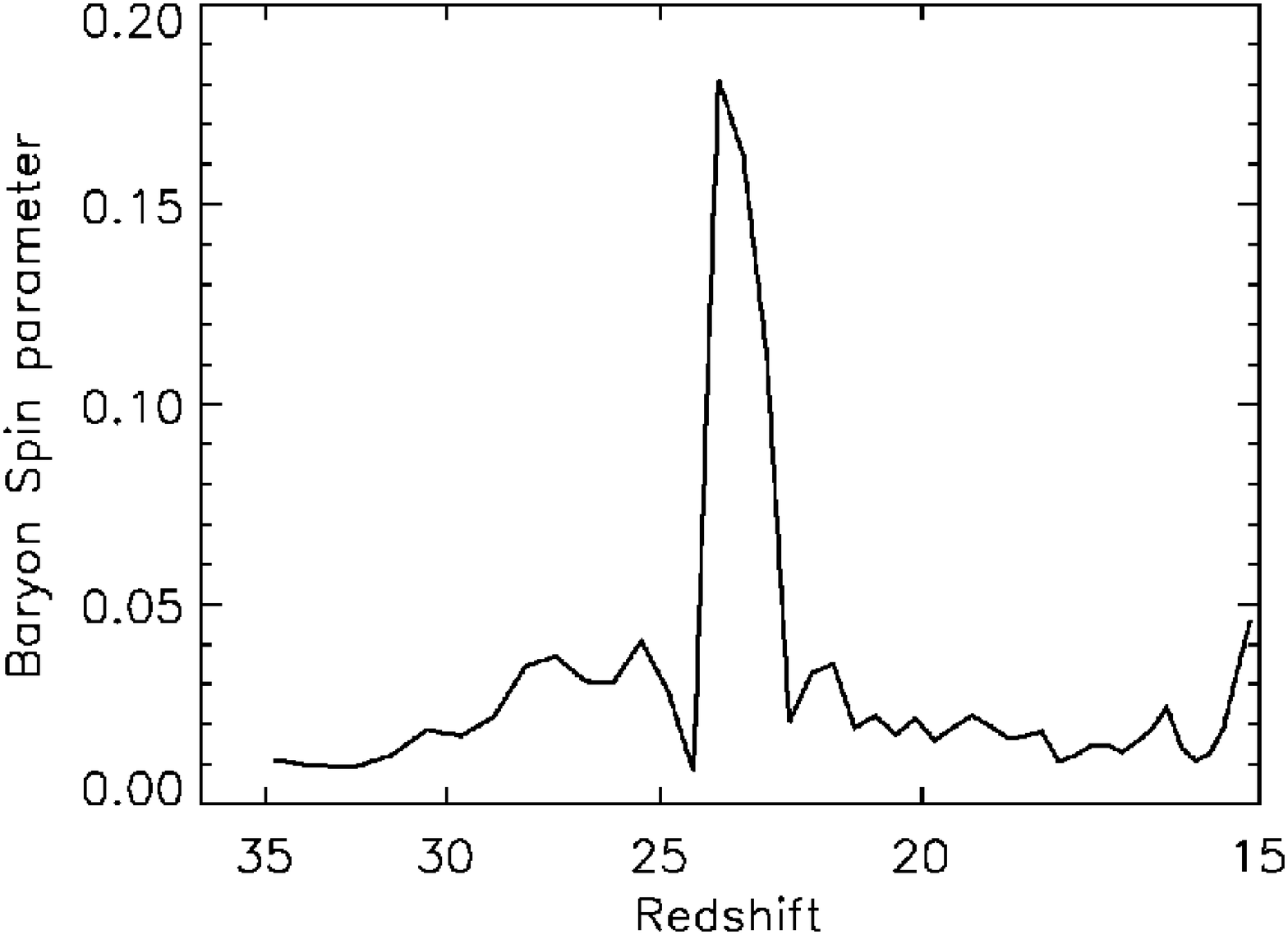}
\caption{Evolution of the spin parameter of the main halo for a metallicity of $Z/Z_{\odot}$=10$^{-4}$.}\label{fig12}
\end{figure}

\begin{figure}
\includegraphics[angle=0,width=8cm]{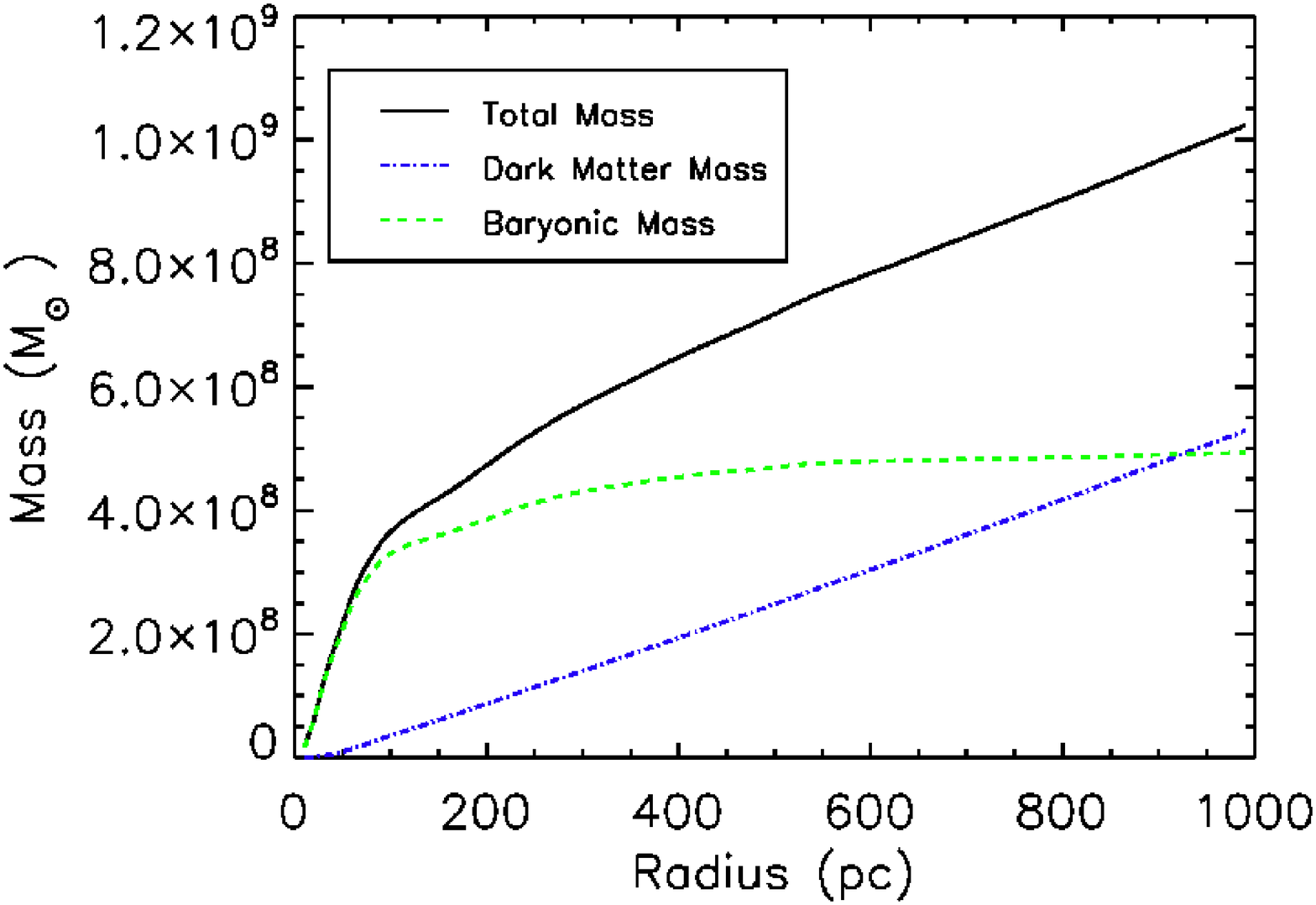}
\includegraphics[angle=0,width=8cm]{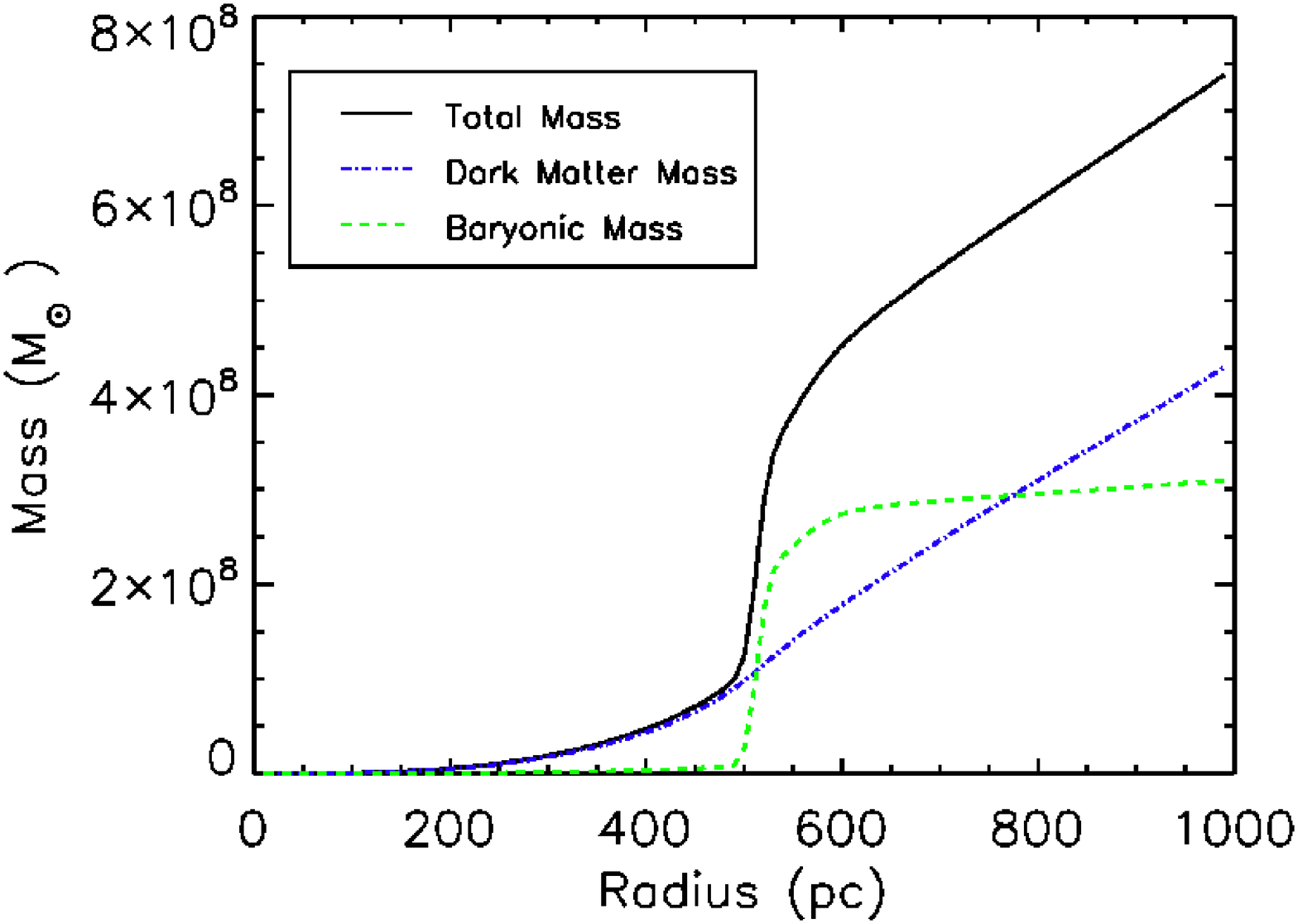}
\caption{Mass vs. radius plots for metallicities of $Z/Z_{\odot}$ = 10$^{-1}$ (top) and  $Z/Z_{\odot}$ = 10$^{-2}$ (bottom) for a radiation field of $G_0$ = 10$^{-2}$ at redshift $z =$ 5. The lines represent the total (black solid),  gas (green dashes), and dark matter (blue dot-dashes).}\label{fig13}
\end{figure}

\begin{figure}
\includegraphics[angle=0,width=8cm]{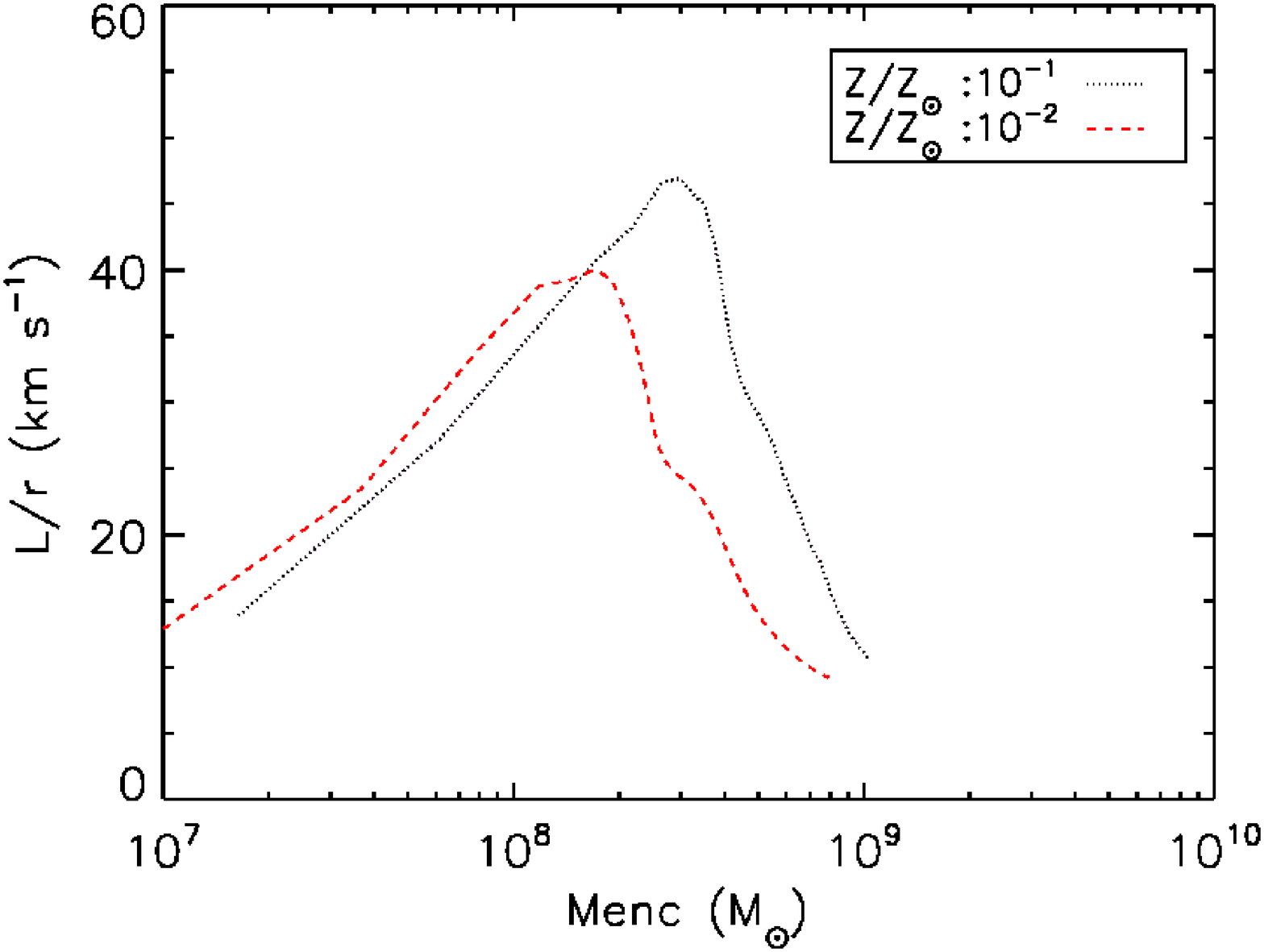}
\includegraphics[angle=0,width=8cm]{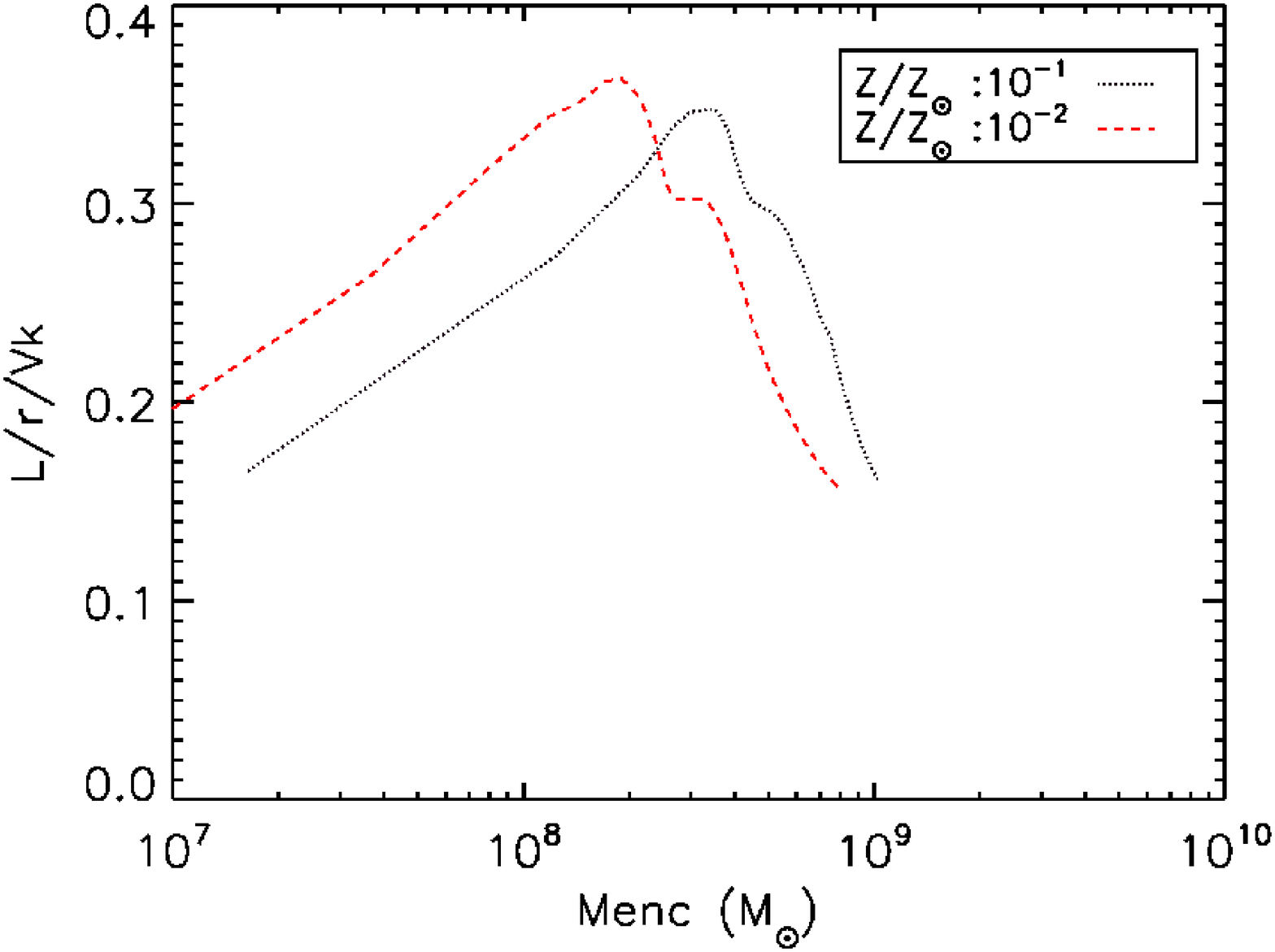}
\caption{Mass-weighted rotational velocity ($L/r$) vs. enclosed gas mass (top) and the ratio of the rotational velocity and circular velocity against enclosed gas mass (bottom) plots for metallicities of $Z/Z_{\odot}$ = 10$^{-2}$ (red dashes) and  $Z/Z_{\odot}$ = 10$^{-1}$ (black dots) at redshift $z =$ 5.}\label{fig14}
\end{figure}

When we look at the density slices of the Z1$-$G1 and Z2$-$G1 runs in Figure \ref{fig3} in the $x-z$ plane we see an elliptical spiral structure and in the $x-y$ plane a flattened structure. This is a disk-like distribution in three dimensions and is more extended and massive in the Z1$-$G1 run ($M =$ 5$\times$10$^9$ $M_\odot$) than in the Z2$-$G1 run ($M =$ 3.4$\times$10$^9$ $M_\odot$). Because the gas cools faster in the halo of the Z1$-$G1 run, more gas is accreted and it settles faster into a disk than in the Z2$-$G1 run.\\
\indent In Figure \ref{fig8}, which is a volume rendering plot, we show the three dimensions density snapshots of the Z1$-$G1 (top), Z2$-$G1 (middle), and Z2$-$G10 (bottom) runs for redshifts $z =$ 7.88 (left), $z =$ 6.24 (middle), and $z =$ 5 (right). We see that in the Z1$-$G1 run more structures are formed and generally they have higher densities than in the Z2$-$G1 and Z2$-$G10 runs. When we  look at the middle and bottom plots we clearly see the effect of UV irradiation on the evolution of a halo. In the Z2$-$G10 run, structures are more fluffy and take more time to become dense and compact due to the heat input. In these plots fluffiness is represented by the spatial extent of the different colors. Although the halo in the Z1$-$G1 run experiences multiple and more violent mergers as well as a higher rate of gas accretion, due to the faster cooling the halo at redshift 5 is denser and more compact than the Z2$-$G1 case. In Figure \ref{fig9}, we plot the amount of mass for the densities indicated along the $x$-axis and for redshifts $z =$ 7.88 (left), z=6.24 (middle), and $z =$ 5 (right).  In these plots it is clearly seen that there is more mass at higher densities in the Z1$-$G1 run than the Z2$-$G1 run. \\
\indent In Figure \ref{fig10}, we plot the fraction of the halo gas mass that has a Jeans mass ($M_J$) below a certain threshold value for the Z1$-$G1 (black$-$dashes) and Z2$-$G1 (red$-$dots) runs. The profile of the low-metallicity case starts with a much steeper rise at low Jeans mass thresholds but then levels off for higher Jeans mass thresholds more rapidly than the high metallicity case. This initial steep rise seems to be due to a lack of low Jeans masses  ($<$ 30 $M_J$) in the central 1 kpc halo of the Z2$-$G1 run. The fact that at any given threshold there is always a larger fraction of gas in the Z1$-$G1 run shows that the Jeans mass is generally lower in the high metallicity case than in the low-metallicity case.\\
\indent As previously mentioned, Figure \ref{fig3} shows a disk-like structure for metallicities $Z/Z_{\odot}$ = 10$^{-2}$ and $Z/Z_{\odot}$ = 10$^{-1}$. In order to see if this disk-like structure is stable we plot the time evolution of specific angular momentum (see Figure \ref{fig11}). In this plot we see that the specific angular momentum increases with time indicating that the disk-like structure becomes more rotationally supported. We also see several peaks which are caused by recent merger events.\\
\indent Next, we plot the spin parameter evolution of our main halo for the Z4$-$G1 run (see Figure \ref{fig12}) . The spin parameter represents the degree of rotational support available in a gravitational system.  We compute the dimensionless spin parameter \citep{1969ApJ...155..393P} as 
\begin{equation}
\lambda \equiv \frac{\rm{|L|} \sqrt{\rm{|E|}}}{GM^{(5/2)}}
\end{equation}
where $L$ is the angular momentum, $E$ is the energy, $M$ is the mass of the object, and $G$ is the gravitational constant. \cite{1987ApJ...319..575B} have calculated that the mean spin parameter is $<\lambda>$ = 0.05 and it has been shown that halos which have suffered a recent major merger tend to have a higher spin parameter than the average (e.g., \cite{2006MNRAS.370.1905H}). As is shown in Figure \ref{fig12}, the spin parameter of our halo peaks around redshift $z =$ 24, where a recent major merger occurred, and overall it has a value of $\approx$ 0.04.\\
\indent In Figure \ref{fig13}, we show the mass radial profiles of the Z1$-$G1 and Z2$-$G1 runs, top and bottom, respectively. In both runs, the inner parts of the halo are dominated by baryonic matter. On the other hand, in the Z2$-$G1 run dark matter takes over at a smaller radius (700 pc) than in the Z1$-$G1 run (950 pc). This is because in the Z1$-$G1 run the halo is larger and more massive baryonically than in the Z2$-$G1 run. Once more, this shows that the halo evolves dynamically slower in the low metallicity run.\\
\indent In Figure \ref{fig14}, we plot the mass-weighted rotational velocity ($L/r$) versus enclosed gas mass (top) and the ratio of rotational velocity to circular velocity versus enclosed gas mass (bottom) for the Z1$-$G1 and Z2$-$G1 runs. In both cases the typical rotational speed is 2$-$3 times lower than the circular velocity which means that the collapse of the halo is not delayed by rotational support \citep{2008ApJ...682..745W}.\\
\subsection{Star Formation and Feedback} 
\begin{figure}
\includegraphics[height=5 cm,width= 7.05cm]{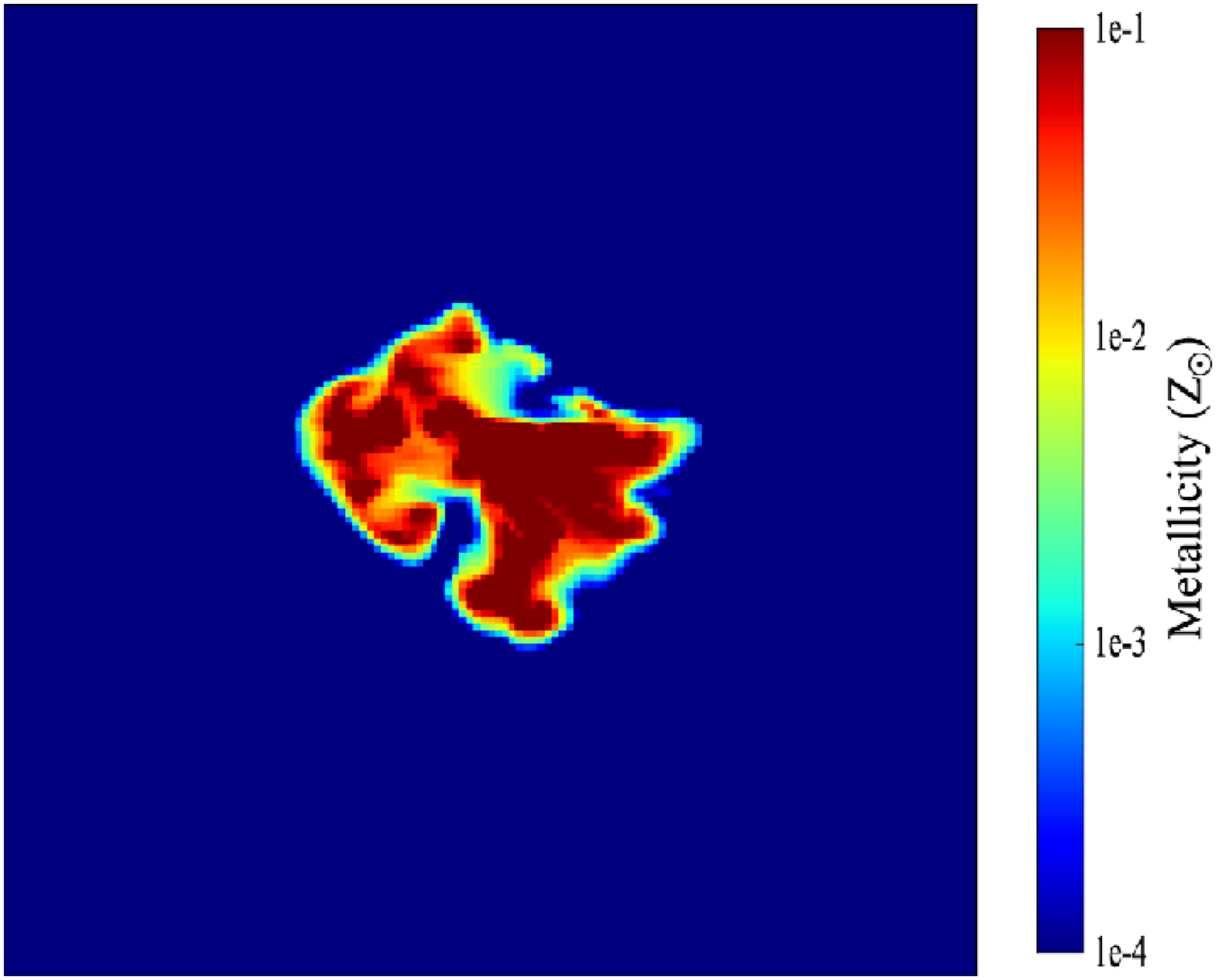}
\vspace{0.05cm}
\includegraphics[height=5 cm,width= 7cm]{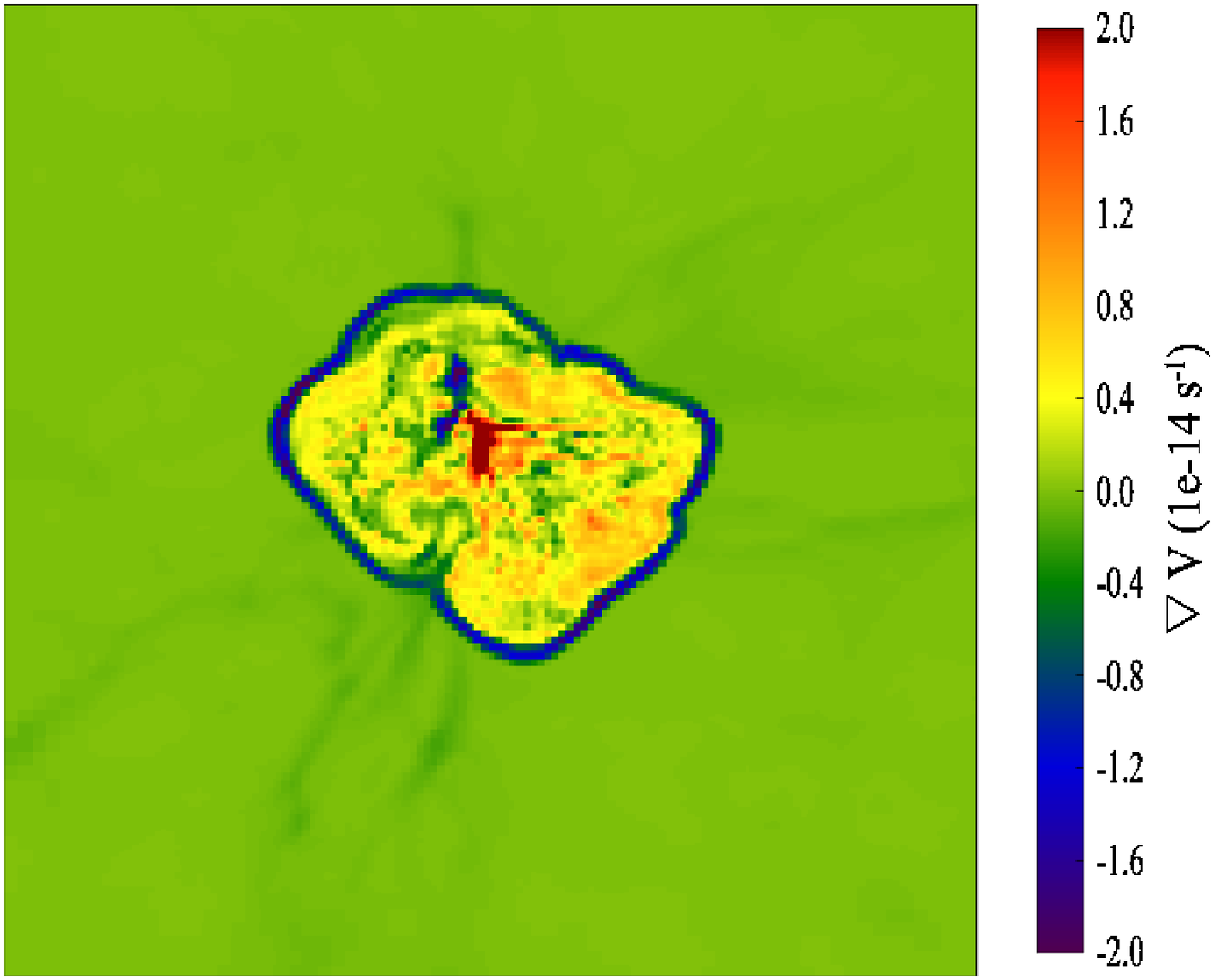}
\caption{Slices of metallicity fraction (top) and velocity divergence (bottom) of the central 10 kpc of a halo for $Z/Z_{\odot}$ = 10$^{-4}$ at redshift 18 in the $y-z$ plane.}\label{fig15}
\end{figure}

\begin{figure*}
\includegraphics[angle=0,width=8cm]{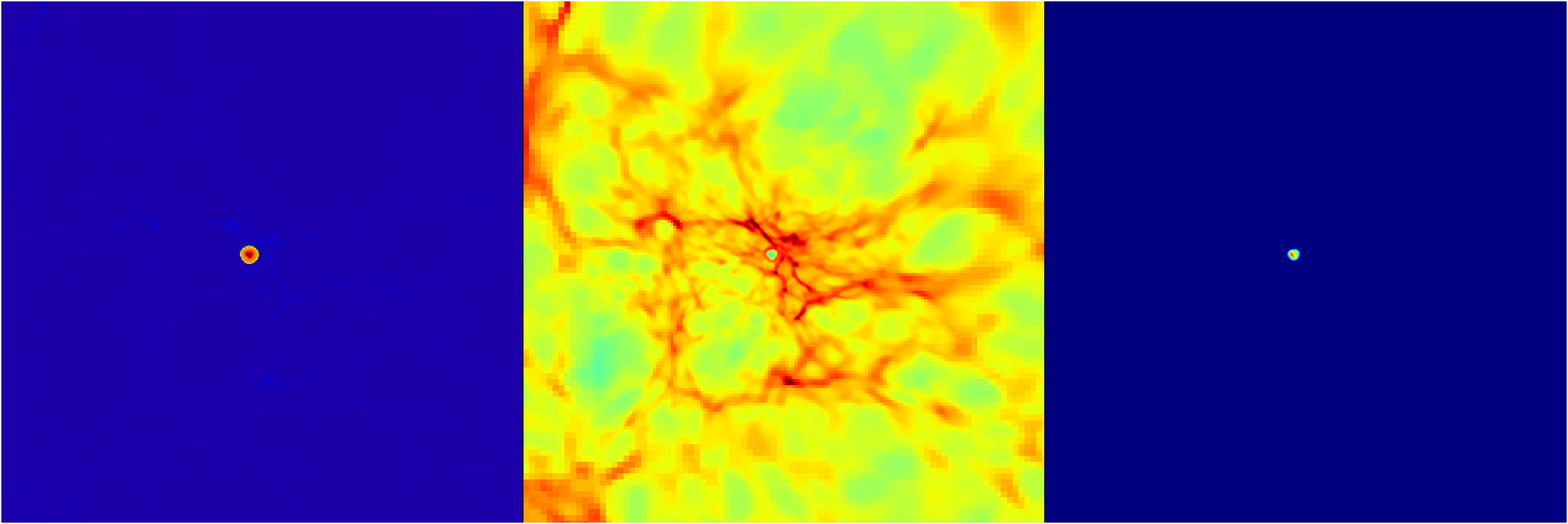}
\vspace{0.05cm}
\includegraphics[angle=0,width=8cm]{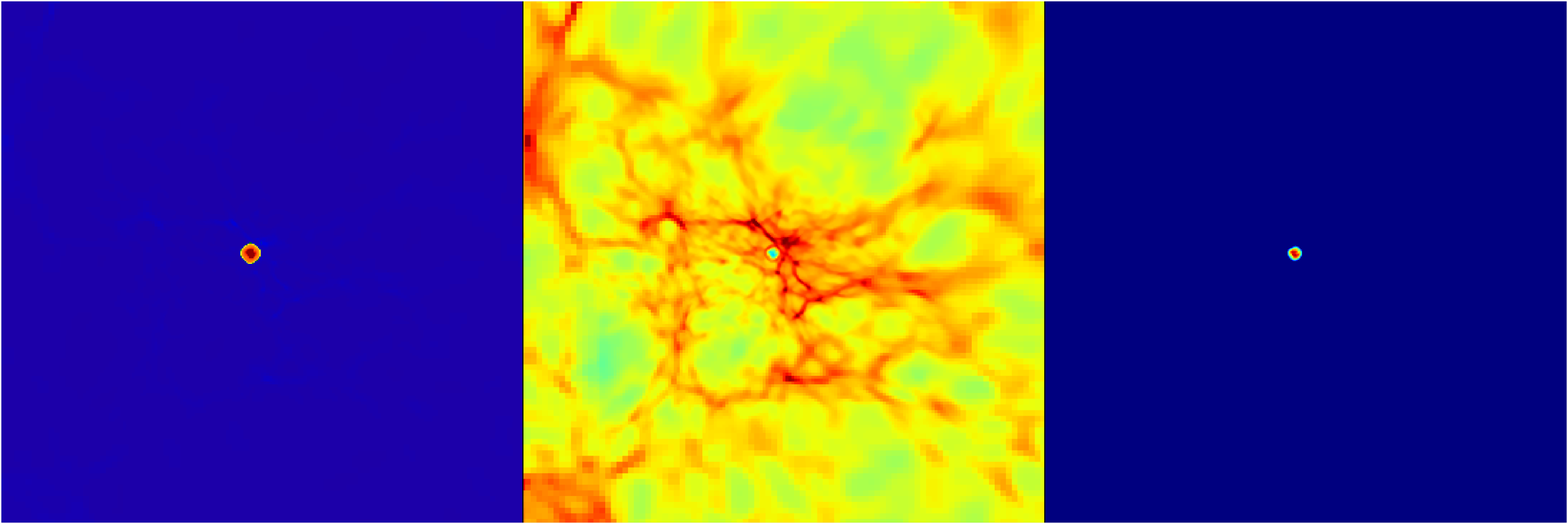}\\
\includegraphics[angle=0,width=8cm]{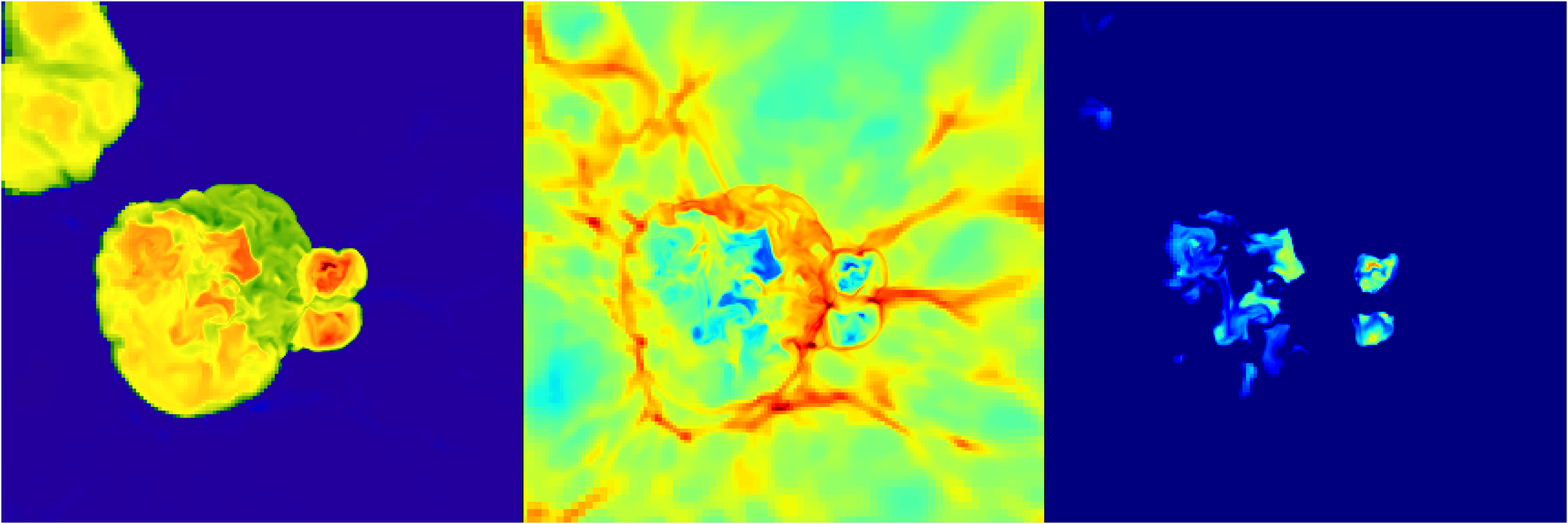}
\vspace{0.05cm}
\includegraphics[angle=0,width=8cm]{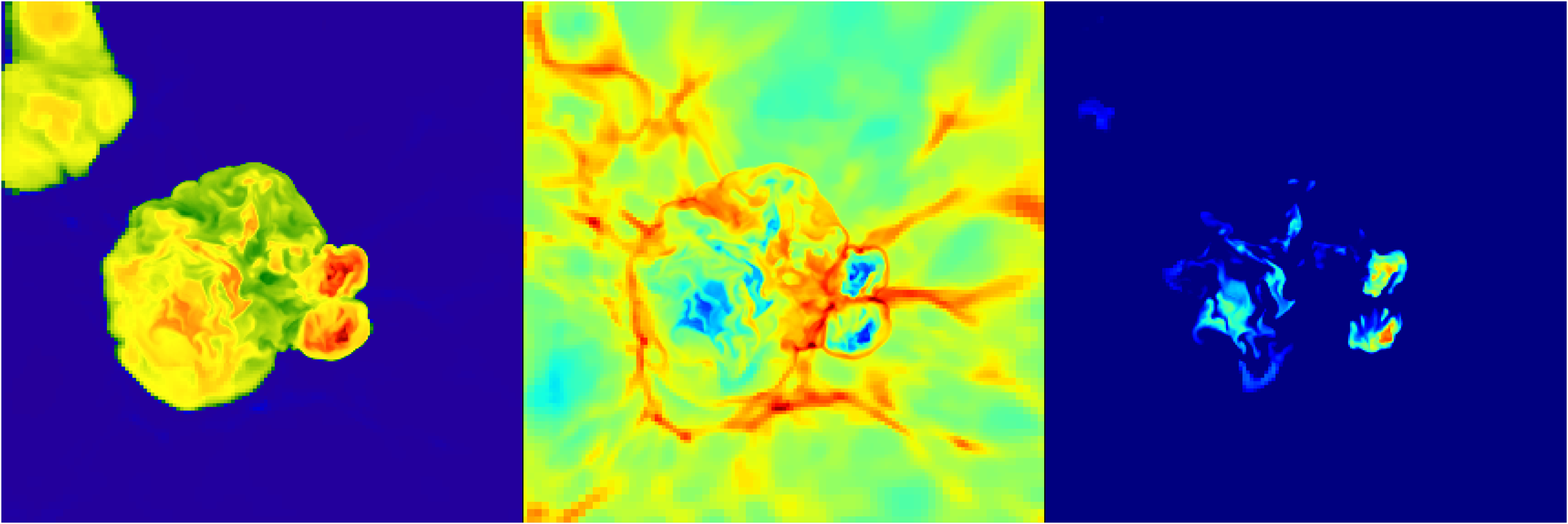}\\
\includegraphics[angle=0,width=8cm]{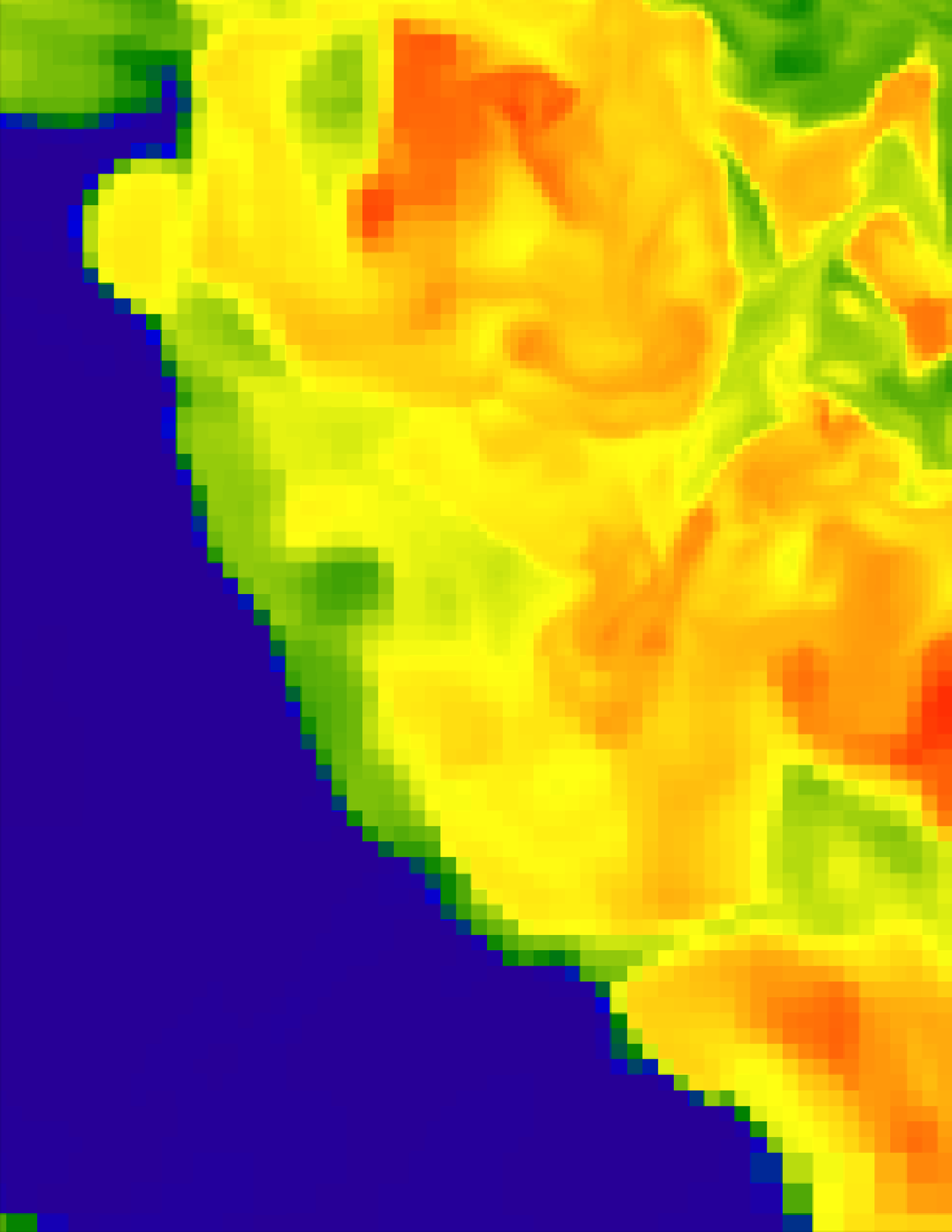}
\vspace{0.05cm}
\includegraphics[angle=0,width=8cm]{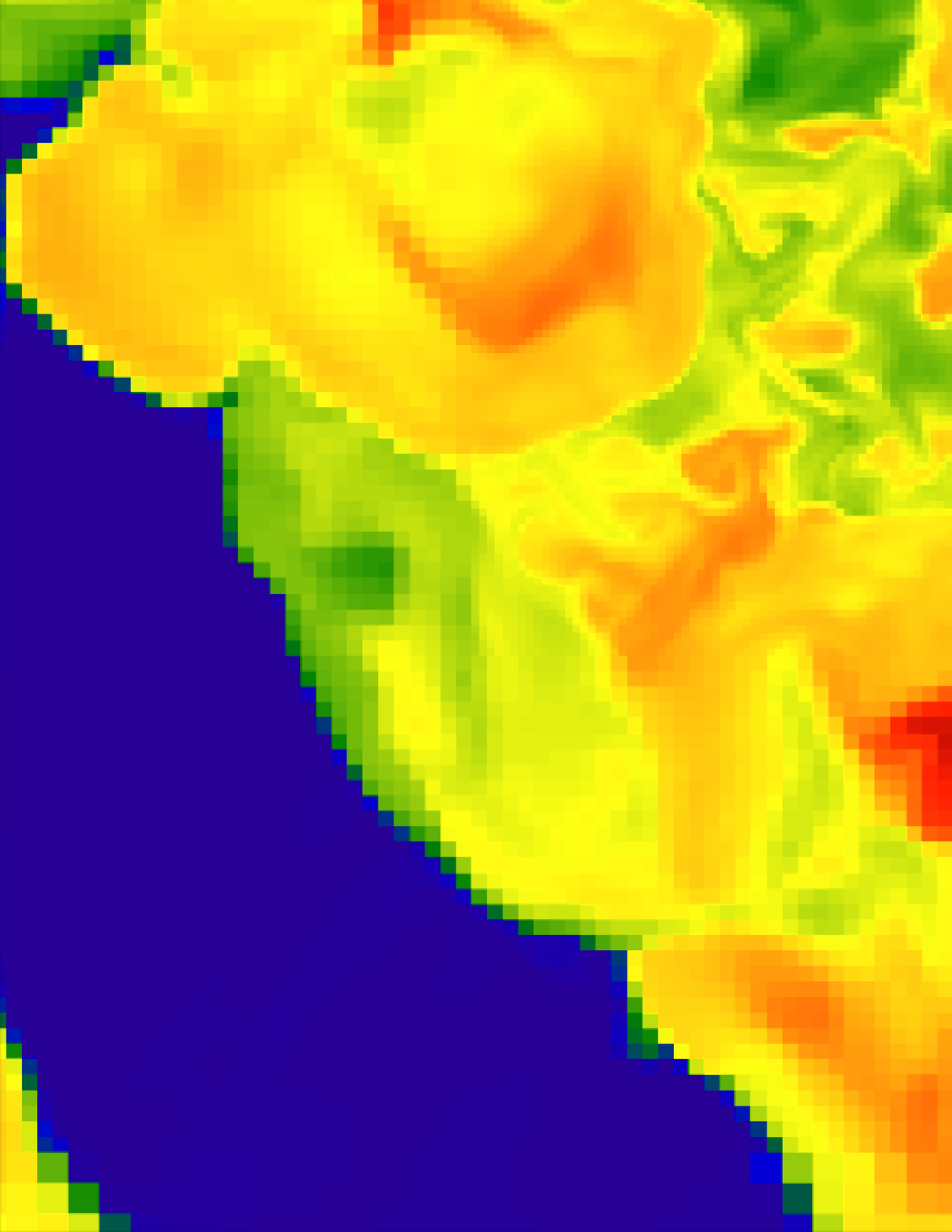}\\
\includegraphics[angle=0,width=8cm]{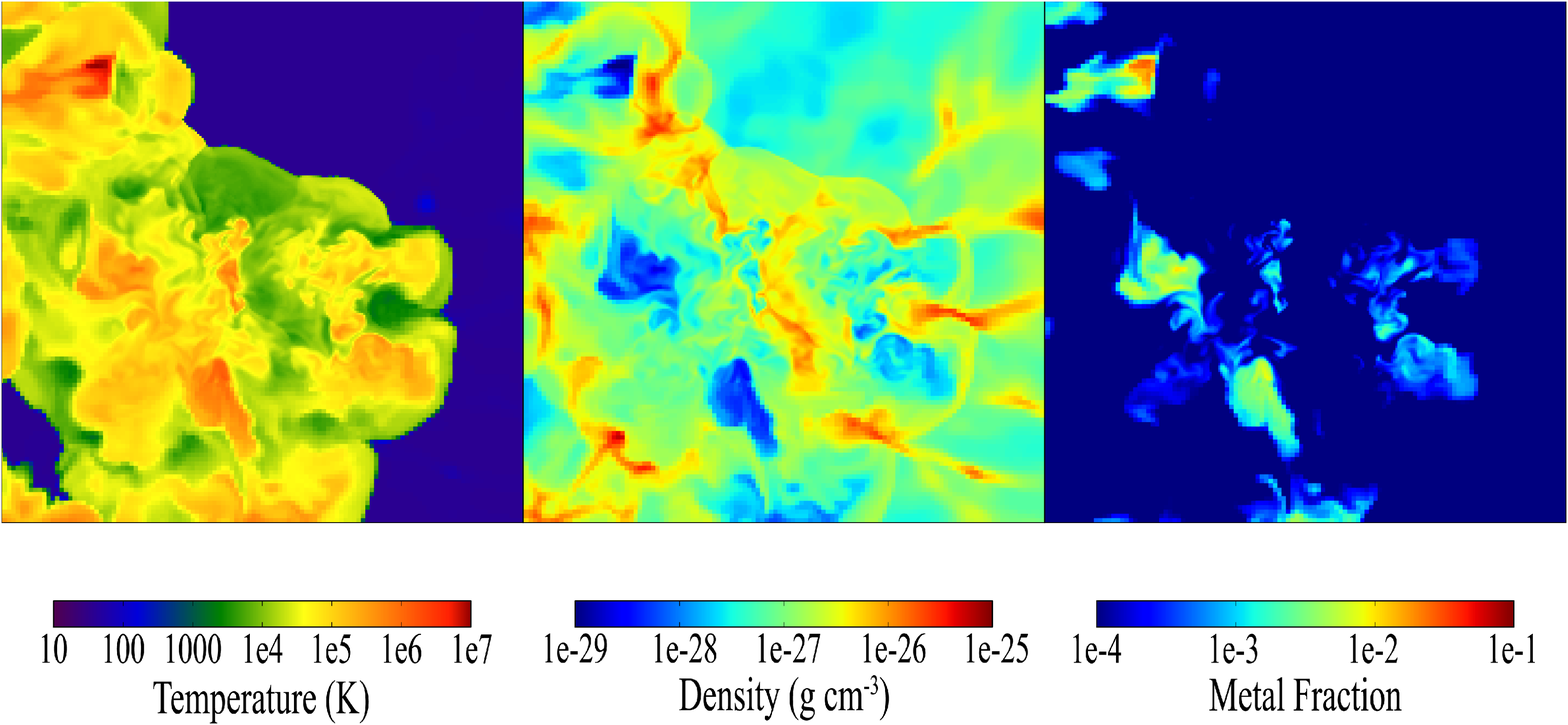}
\vspace{0.05cm}
\includegraphics[angle=0,width=8cm]{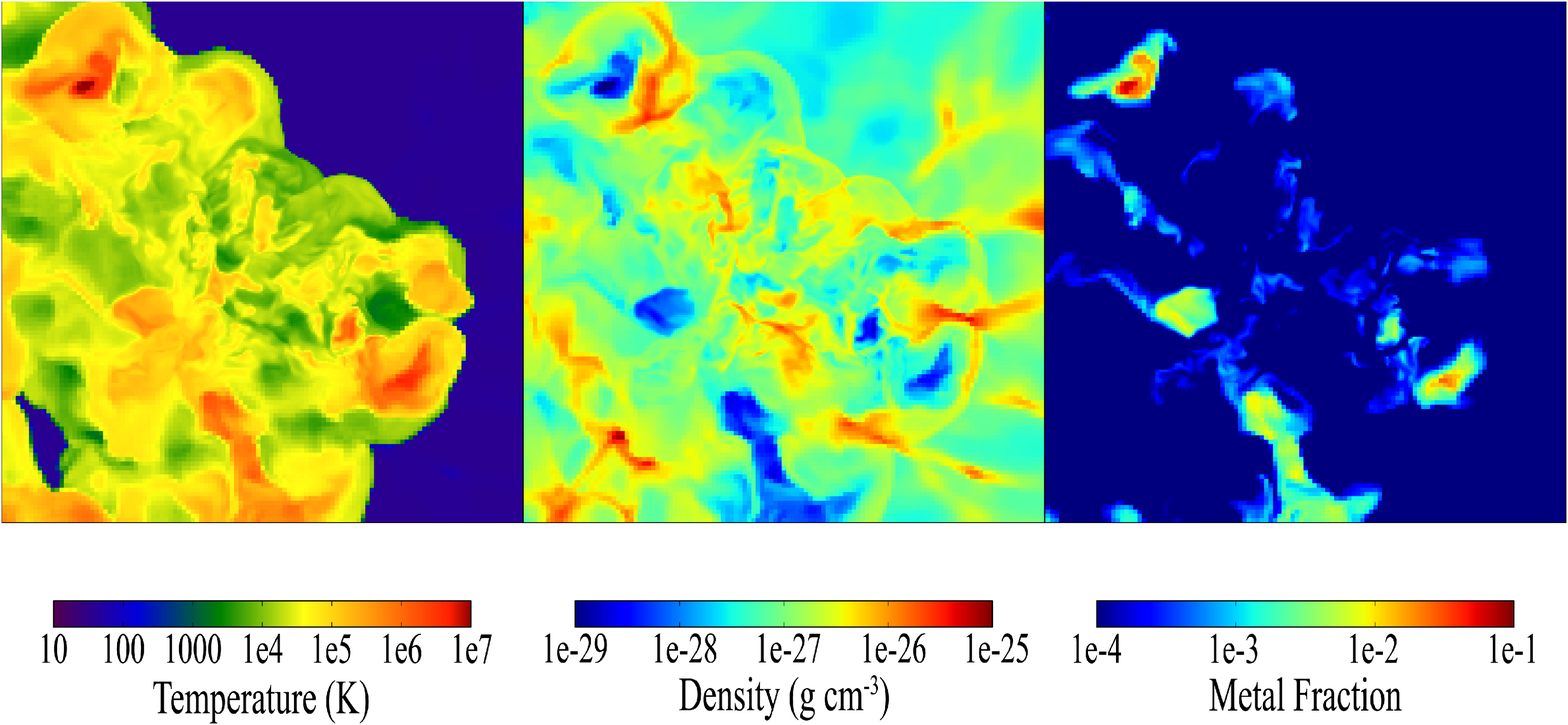}
\caption{From left to right: Slices of temperature, density, and metallicity fraction of the central 25 kpc for pre-enrichment metallicities of $Z/Z_{\odot}$ = 10$^{-4}$ (left column) and $Z/Z_{\odot}$ = 10$^{-2}$ (right column) for a radiation field of $G_0$ = 10$^{-2}$ at redshifts $z =$ 18.7, 15, 13.4, and 12.3 (from top to bottom, respectively) in the $y-z$ plane.} \label{fig16}
\end{figure*}

When we include star formation and mechanical feedback in our simulations we see that, in the case of $Z/Z_{\odot}$ = 10$^{-4}$ pre-enrichment, the halo is further enriched up to  $Z/Z_{\odot}$ = 10$^{-1}$ by the SNe of the first stars as early as $z \sim$18. In Figure \ref{fig15}, we present the metal fraction (top) and the velocity divergence (bottom) of the central 10 kpc around the halo a short time after the first Pop III SN exploded. In the velocity divergence plot, we see a clear ring-like structure with a large velocity gradient. This is due to the multiple outflows driven by SNe that collide with inflowing material. Also, the highest outflow velocities are right behind the shock front as expected.\\
\indent In Figure \ref{fig16}, where we plot time snapshots of temperature, density, and metallicity fraction of the central 25 kpc right after the first SNe, it is seen that the gas is heated to $\geq$ 10$^4$ K and expelled from the halo and hence infall of the outer halo gas onto the central core of the halo is quenched. The distribution of metals, provided by SNe, is very patchy. In the center of the halo, there are regions present with high metallicities, $Z/Z_{\odot}$ = 0.1, while in the outskirts of the halo the gas is still poor in metals. This indicates that SN feedback efficiently suppresses star formation in our enriched halos, irrespective of the ambient UV field. \\
\indent On the other hand, after the first SNe go off we do not see any significant difference in the distribution of metals in the halo for the pre-enriched runs with $Z/Z_{\odot}$ = 10$^{-4}$ and $Z/Z_{\odot}$ = 10$^{-2}$, as shown in Figure \ref{fig16}. This indicates that the higher densities in the high-metallicity run do not have any effect on the mixing of metals in the halo. There might be a cooling instability effect on the mixing of the metals produced by the SNe with the surroundings that we are not taking into account, but since the SN feedback is strong we think that this effect is negligible.\\
\section{CONCLUSIONS AND CAVEATS}\label{Fin}
\indent We summarize our results as follows.
\begin{enumerate}[1)]
\item The redshift at which a multi-phase ISM is established depends on metallicity.  In the absence of UV radiation, this critical metallicity is consistent with \cite{2003Natur.425..812B} and \cite{2007ApJ...661L...5S}, and is ($Z/Z_{\odot}$)$_{crit}$ $\sim$ 10$^{-3.5}$.
\item Above a metallicity of 1$\%$ Solar the cooling efficiency of ambient gas no longer increases with a rise in metallicity \citep{2008ApJ...678L...5S}, and therefore the Jeans masses in halos that are pre-enriched to metallicities of $Z/Z_{\odot}$ = 10$^{-2}$ and $Z/Z_{\odot}$ = 10$^{-1}$ are comparable.\\
\item The cold dense gas phase is fragile to UV radiation for metallicities of $Z/Z_{\odot}$ $\leq$ 10$^{-3}$ and robust to UV radiation for metallicities of $Z/Z_{\odot}$ $\geq$ 10$^{-2}$. Thus, the metal-poor star-forming ISM is fragile to UV radiation, and inclusion of a constant radiation background raises the critical metallicity value for the Pop III$-$Pop II transition from Z$_{cr}$ $\sim$ 10$^{-3.5}$$Z_{\odot}$ to $Z_{cr}$ $\sim$ 10$^{-2}$$Z_{\odot}$ when $F_0$ $>$ 10$^{-5}$ erg s$^{-1}$ cm$^{-2}$, which is in good agreement with the values that are found for the suppression of H$_2$ by \cite{2010MNRAS.402.1249S}. This is because H$_2$ is one of the main drivers of ion$-$molecule chemistry in metal-enriched gas and without it cooling is dominated by the fine-structure lines of [C II] and [O I] rather than CO. 
\item All the indicators for the dynamical evolution show that the halo evolves dynamically faster for $Z/Z_{\odot}$ $\geq$ 10$^{-2}$  and that the cold and dense gas phase in high metallicity halos survives (violent) mergers due to the enhanced cooling.
\item Pre-enrichment does not affect mixing of the metals that are produced by the first SNe.
\end{enumerate}
\indent For future work, our simulations can be improved in several ways. We reach down to a maximum resolution of 75 and 7 pc with and without star formation feedback in our simulations, respectively. This is not enough to resolve the smallest gas fragments that will eventually turn into stars. Our simulations also assume a constant ambient UV background field which might not be realistic because the UV radiation will originate from specific stellar sources. These sources are most likely to form in high-density areas and it would therefore be better to tie the UV background radiation to these sources at the time of their formation. \\
\indent Another improvement can be made by adjusting the way we pre-enrich the simulations. Currently we assume one metallicity for a specific simulation. We do this so that we can confidently study specifically the metallicity effects on the evolution of our halo. However, the distribution of metals, that are formed by the first SNe, is not uniform but inhomogeneous. Therefore, in the near future we will perform simulations that consider patchy pre-enriched halos.



\acknowledgments
We acknowledge the computational resources of the University of Groningen, the millipede cluster, and the Gemini machines at the Kapteyn Astronomical Institute. Computations described in this work were performed using the Enzo code developed by the Laboratory for Computational Astrophysics at the University of California in San Diego (http://lca.ucsd.edu). We are grateful for insightful comments
from an anonymous referee. A.A. acknowledges P. Kamphuis for useful discussions and support.






\bibliographystyle{apj}   
\bibliography{references}

\end{document}